\documentclass[prb,aps,amssymb,amsmath,superscriptaddress,twocolumn,showpacs]{revtex4-1}

\usepackage{graphicx}
\usepackage{dcolumn}
\usepackage{bm}
\usepackage{epstopdf, color}

\let\a=\alpha \let\b=\beta    \let\g=\gamma     \let\d=\delta     \let\e=\varepsilon
  \let\h=\eta           
\let\m=\mu    \let\n=\nu              \let\p=\pi        \let\r=\rho
\let\s=\sigma             \let\c=\chi
   \let\o=\omega     
\let\G=\Gamma \let\D=\Delta       \let\L=\Lambda    
             
\let\O=\Omega

\def\be{\begin{equation}}
\def\ee{\end{equation}}
\def\bea{\begin{eqnarray}}\def\eea{\end{eqnarray}}
\def\bean{\begin{eqnarray*}}\def\eean{\end{eqnarray*}}

\def\nn{\nonumber}

\def\WW{{\cal W}}
 
\def\BBB{{\cal B}}
\def\RR{{\cal R}}\def\LL{{\cal L}} 

\def\qq{{\bf q}} \def\pp{{\bf p}}
\def\xx{{\bf x}} \def\yy{{\bf y}} \def\zz{{\bf z}}
\def\kk{{\bf k}}

\def\media#1{{\langle#1\rangle}}
\def\mmedia#1{{\langle\!\!\langle#1\rangle\!\!\rangle}_{\!\infty}}

\def\Tr{\mathrm{Tr}}

\begin{document}

\title{Topological phase transitions and universality in the Haldane-Hubbard model}
\author{Alessandro Giuliani}
\affiliation{University of Roma Tre, L.go S. L. Murialdo 1, 00146 Rome, Italy}
\author{Ian Jauslin}
\affiliation{University of Z\"urich, Winterthurerstrasse 190, 8057 Z\"urich, Switzerland}
\author{Vieri Mastropietro}%
\affiliation{University of Milan, Via Saldini 50, Milan, Italy}

\author{Marcello Porta}
\affiliation{University of Z\"urich, Winterthurerstrasse 190, 8057 Z\"urich, Switzerland}
\begin{abstract} 
We study the Haldane-Hubbard model by exact Renormalization Group techniques. We analytically construct the topological phase diagram, for weak interactions. We predict that many-body interactions induce a shift of the transition line: in particular, repulsive interactions enlarge the topologically non-trivial region. 
The presence of new intermediate phases, absent in the non interacting case, is rigorously excluded at weak coupling. Despite the non-trivial renormalization of the wave function and of the Fermi velocity, the conductivity is universal: at the renormalized critical line, both the discontinuity of the transverse conductivity and the longitudinal conductivity are independent of the interaction, thanks to remarkable cancellations due to lattice Ward Identities. In contrast to the quantization of the transverse conductivity, the universality of the longitudinal conductivity cannot be explained via topological arguments.
\end{abstract}

\maketitle


\section{Introduction}

The current understanding of topological matter \cite{K,K1,K2} is mostly based on a single-particle description. A paradigmatic example is the integer quantum Hall effect: in the absence of interactions, the Hall 
conductivity has a deep topological interpretation\cite{ASS,ASS1}, which explains its quantization and stability. A more recent example is provided by the classification of time-reversal invariant insulators
\cite{KM,KM1,KM2,KM3,KM4,KM5,KM6,KM7}, 
which, again, relies on the properties of the noninteracting Bloch functions. Understanding the effect of interactions on topological matter has become a very active area of research \cite{H0}. 

A natural model in which to explore such issues is the {\it Haldane-Hubbard model}. The Haldane-Hubbard model describes spin-1/2 electrons on the honeycomb lattice, interacting via a local Hubbard interaction of 
strength $U$. The electrons hop between nearest neighbor sites with hopping strength $t_{1}$, and between next-to-nearest neighbor sites with alternating hopping parameters $t_{2}e^{\pm i\phi}$: the phases 
$\pm\phi$ describe a transverse magnetic field, with zero net flux through the honeycomb plaquette. Finally, the system is also exposed to a staggered chemical potential, with strength $\pm W$ on the two 
triangular sublattices.  In the absence of interactions \cite{H} this model shows, depending on the value of its parameters, a trivial insulating phase with vanishing transverse conductivity $\s_{12} = 0$, or a 
quantum Hall phase with $\s_{12}= \pm2\frac{e^2}{h}$. These topological phases are separated by two critical curves in the $(\phi, W)$ plane, intersecting at the crossing points $(0,0)$ and $(\pi,0)$. Along the 
critical curves, the energy bands touch at a conical intersection; at the crossing points, there are two such conical intersections, as in standard graphene. Indeed if $t_2=W=0$ the system describes 
graphene with short range interaction.

From a theoretical viewpoint, the Haldane topological phases have been argued to emerge in pure graphene sheets by spontaneous mass generation, due to the strong, unscreened Coulomb repulsion 
\cite{H1,H11, H12,H13,H14,H15,H16, GMPgauge}. From an experimental viewpoint, the Haldane model has been realized in Ref.\cite{E}, and the topological phase transition has been observed. 
The inclusion of a tunable Hubbard interaction seems
to be accessible by the present technology. Therefore, studying its effects on the transport coefficients is of fundamental importance for the next generation of cold atom experiments. 
So far, the properties of the Haldane-Hubbard model have been investigated mostly via mean-field, variational and numerical analyses\cite{H3,H31,H32,H33,H34,H35,H36,H37,H38,H39,H310,H311,H312,
 Van, Hub}. 
 
Concerning the {\it transverse conductivity}, topological arguments for interacting systems\cite{AS,HM} ensure that, away from the critical curves, $\s_{12}$ can only take integer values, in units of $e^{2}/h$ (here 
$\s_{ij}$ are the 
elements of the conductivity matrix, in the limit of zero frequency and zero temperature). However, its specific value at a given point in the phase diagram
can be different from the corresponding non-interacting value, in particular in the vicinity of the critical curves 
(at weak coupling, far from the critical lines,  the conductivity is known to be independent of the interaction 
\cite{CH, GMPhall}). The relevant question here is to distinguish between two scenarios: the first, in which small interactions are not able to generate new phases and their main effect is a shift of the critical curves, 
as found in certain 3D topological insulators \cite{E2,E21,E22}; and a second one, characterized by 
the emergence of a novel, interaction-induced, topological phase, like the one corresponding to $\s_{12} = \pm \frac{e^{2}}{h}$, predicted for the Haldane-Hubbard model 
in Ref.\cite{H3,H31,H32,H33,Van, Hub}. Regarding the {\it longitudinal conductivity}, in the absence of interactions it is equal to $\frac{e^2}{h} \frac{\pi}{4}$, for all the values of $(\phi,W)$ {\it on} the critical lines, 
with the exception of the crossing points $(\phi,W)=(0,0),(\pi,0)$,
where it is equal to $\frac{e^2}{h} \frac{\pi}{2}$ (of course, away from the critical lines $\s_{ii}=0$). There are no topological arguments ensuring that the {\it critical} longitudinal conductivity 
should remain quantized when the interaction is switched on: therefore, the relevant question here is whether the interaction introduces corrections breaking this exact quantization or not. 
This question is related to a similar one discussed in the context of graphene, 
in which recent experiments \cite{Nair} showed that the optical longitudinal conductivity is essentially universal, and in excellent agreement with the value computed for the non-interacting model \cite{SPG}; on 
the contrary, the interaction produces
dramatic effects on other physical quantities, such as the Fermi velocity \cite{Elias}. 
On the theoretical side, the universal behavior in graphene is still not completely understood \cite{L1,L11,L12,L13,L14,L15,L16,L17,L18,L19}, see, in particular, Ref.\cite{L17} for a recent review. 
A rigorous result for short range interaction 
\cite{GMP} showed that, in order to get exact universality of graphene's longitudinal conductivity, one needs to fully take into account the non linear correction to the bands, even if such terms are irrelevant in 
the Renormalization Group (RG) sense.

In this paper we compute the conductivity matrix of the Haldane-Hubbard model via exact RG methods, close to and at the critical lines, for weak interactions. 
We take lattice effects into account, and we exploit {\it lattice} symmetries in 
order to reduce the number of independent running couplings, in a way similar to Ref.\cite{GM, GMP, FA,HJR} for graphene. 
The use of exact RG methods is motivated by the fact 
that the computation of conductivity is extremely sensitive to the choice of the regularization scheme\cite{L19}. 
Even though they are irrelevant in the RG sense, lattice and interaction effects produce, in general, {\it finite corrections} to the physical observables, and they must be taken into account 
in order to prove or disprove the emergence of new interaction-induced topological phases, as well as to address the issue of universality of the critical longitudinal conductivity. 

\begin{figure}[ht]
\includegraphics[width=.5\textwidth]{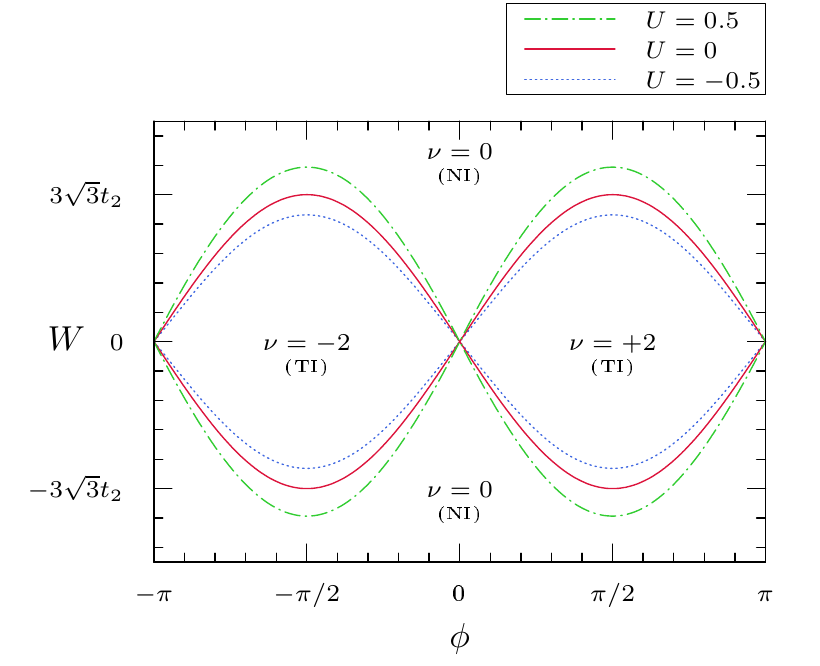}
\caption{Interacting phase diagram of the Haldane-Hubbard model for $t_1=1$, $t_2=0.1$, and different values of $U$. $\s_{12} = (e^{2}/h)\nu$, where the values of $\nu$ are reported in the figure.
for $\n=\pm2$ the system is a topological insulator (TI),
while for $\n=0$ the system is a trivial, normal, insulator (NI).}
\label{figaltezza}
\end{figure}
\begin{figure}[ht]
\includegraphics[width=.5
\textwidth]{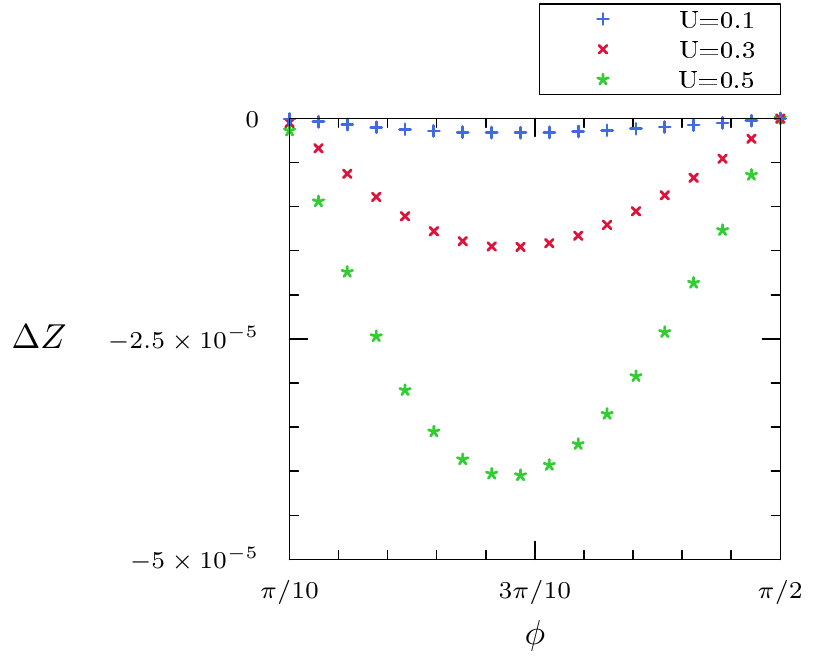}
\caption{The difference of the wave function renormalizations $\Delta Z=Z_{1,R}-Z_{2,R}$ on the critical line, as a function of $\phi$, for different values of $U$. This difference would be zero for the effective relativistic theory.}
\label{fig1-2}
\end{figure}

By choosing the chemical potential $\mu\equiv \mu(U)$ so to fix the Fermi energy half-way between the valence and conduction bands, the band gap can only close at the two Fermi points $\vec p_F^{\,\o}=(\frac{2\p}3,\o\frac{2\p}{3\sqrt3})$, where $\omega = \pm$ is the {\it valley} 
index. We prove that, close to criticality, the interacting Euclidean two-point function is:
\bea &&\nn
\hat S_2(k_0,\vec p_F^{\,\o}+\vec k') =\\
&&=-\begin{pmatrix} ik_0 Z_{1,R}-m_{R}& v_R(-ik_1'+\o k_2')\\
v_R(ik_1'+\o k_2') & ik_0 Z_{2,R}+m_{R}\end{pmatrix}^{-1}\hskip-.1truecm\big(1+R(k_0,\vec k')\big)\nn\eea
where the error term $R(k_0,\vec k')$ is subleading in the effective mass $m_R$, in the Matsubara frequency $k_0$ and in the quasi-momentum $\vec k'$.
The parameters $Z_{1,R},Z_{2,R},v_R, m_R$, depend non-trivially on the valley index $\omega$ and on the interaction. In particular, 
the renormalized mass $m_{R}\equiv m_{R,\o}$ reads:
\be\label{fpm}
m_{R,\pm}=W \pm 3\sqrt{3}\,t_{2}\sin\phi-F_{\pm}(U,W,\phi)
\ee
where $\pm$ is the valley index, and $F_\pm$ is expressed in the form of a {\it convergent} renormalized series, whose first non-trivial order is given by Eq.\eqref{eq.Fnumeric} below.
The dressed critical lines, defined by the condition that the renormalized mass vanishes, are also modified
by the interaction, see Fig.\ref{figaltezza}. Similarly, the Fermi velocity $v_R$ and the wave function renormalizations $Z_{1,R}, Z_{2,R}$
have non-trivial interaction corrections and, remarkably, $Z_{1,R}$ and $Z_{2,R}$ are {\it different}, as shown in Fig.\ref{fig1-2}. All these 
non-universal renormalizations are {\it absent} in effective relativistic descriptions: 
by neglecting the (irrelevant) non-linear corrections to the energy bands, one would obtain a Nambu-Jona Lasinio model, 
in which Lorentz and chiral symmetry would imply the invariance of $v_R$, the invariance of the critical lines and 
$Z_{1,R}= Z_{2,R}$. However, these extra symmetries are broken by the lattice, and the renormalization of the effective parameters
is a physical signature of many body interaction that should 
be visible in real systems, e.g., in cold atom experiments. These non-universal parameters also enter 
the computation of the conductivity: remarkably, they are related by exact {\it lattice Ward identities},
which induce non-trivial cancellations and imply subtle universality properties, 
as stated in the following theorem.
We recall that $\s_{ij}$ are the 
elements of the Kubo conductivity matrix, in the limit of zero frequency and zero temperature. We also denote by $\s
_{ij}^{cr}$ their values on the renormalized critical curves. 

\medskip

\noindent{\bf Theorem.} {\it There exists $U_{0}>0$ such that for $-U_{0} < U < U_{0}$, the system is massless if and only if the right side of Eq.\eqref{fpm} vanishes.
This condition defines two renormalized critical curves intersecting at $(\phi,W)=(0,0),(\pi,0)$, separating two non-trivial topological phases, characterized by transverse conductivity 
$\s_{12}=\pm2(e^2/h)$, from two standard insulating phases, see Fig.\ref{figaltezza}. On the renormalized critical curves, the critical longitudinal conductivity $\s_{ii}^{cr}$, $i=1,2$, is quantized: 
if $\phi\neq0,\pi$, 
\be
\s_{ii}^{\text{cr}} = \frac{e^2}{h} \frac{\pi}{4}\;,
\ee
while $\s_{ii}^{\text{cr}} = \frac{e^2}{h} \frac{\pi}{2}$ at $(\phi,W)=(0,0),(\pi,0)$.
}

\medskip

Thus, the critical lines acquire non-universal, interaction-dependent corrections, but they still separate topological regions labelled by $\nu = \pm 2$ from the trivial ones, labelled by $\nu = 0$, 
see Fig.\ref{figaltezza}.
New intermediate phases characterized by the quantum number $\n=\pm1$ are {\it rigorously excluded} at weak coupling: the universality class of the topological transition remains unchanged. 
The effect of the repulsive interaction is to {\it enlarge} the topologically non-trivial region, see Fig.\ref{figaltezza}. This enhancement agrees with the numerical findings of Ref.\cite{Van,Hub} and
is presumably a sign that repulsive interactions in graphene-like systems can favor the spontaneous generation of the topological insulating phase \cite{H1,H11, H12,H13,H14,H15,H16, GMPgauge}.
 
Even if not protected by any topological argument, the critical longitudinal conductivity $\s_{11}^{\text{cr}}$ is {\it exactly universal} and equal to half the 
one of graphene, on the whole critical line, with the exception of the special crossing points $(0,0)$ and $(\pi,0)$, at which the value of $\s_{11}^{\text{cr}}$ is the same obtained for interacting graphene\cite{GMP}, namely $(e^{2}/h)(\pi/2)$: each Dirac cone contributes with a universal quantity $(e^{2}/h)(\pi/4)$ to the critical longitudinal conductivity. Of course, away from the critical curves, 
the longitudinal conductivity is exactly zero. 

Our results are in agreement with a low energy description in terms of an effective action that includes a non-trivial Chern-Simons term, whose coefficient (the Hall conductivity) is proportional to the difference of the signs of the renormalized masses, $\text{sign} (m_{R,-}) - \text{sign} (m_{R,+})$, rather than the bare ones, as one would get in the relativisitic approximation \cite{LYL}.

The theory that we develop is {\it non perturbative}, in the sense that it allows us to express all the correlations and transport coefficients in terms of {\it convergent} series. 
As it will appear from the analysis, our non-perturbative bounds on the correlation functions, once combined with Ward Identities, allow us to conclude the universality of the conductivity, 
without exploiting explicit cancellations at all orders. 
We have not tried to optimize the estimate for the radius of the convergence domain, which, therefore, is expected to be far from the values of $U$ where interaction-induced phase transitions might take place. 
However, we believe that the range of validity of our convergent expansions could be improved by combining our analysis with numerical techniques, as it is done, for instance, for the stability of KAM tori in 
classical mechanics. Finally, we stress that our analysis only requires the interaction to be short-ranged, we considered the Hubbard interaction just for the sake of definiteness.

The paper is organized as follows. In Section \ref{sec2} we define the Haldane-Hubbard model, and derive the exact lattice Ward Identities for its correlation functions. 
In Section \ref{sec3} we perform an exact Renormalization Group analysis for the correlations, we classify the allowed running coupling constants by the exact lattice symmetries of the system, 
and we compute the decay of the correlations at large distances, as well as the renormalized critical line. 
In Section \ref{sec4} we prove the quantization of the Hall conductivity across the critical line, and the universality of the critical longitudinal conductivity.

\section{The Haldane-Hubbard model}\label{sec2}

\begin{figure}[hbtp]
\centering
\includegraphics[width=.3
\textwidth]{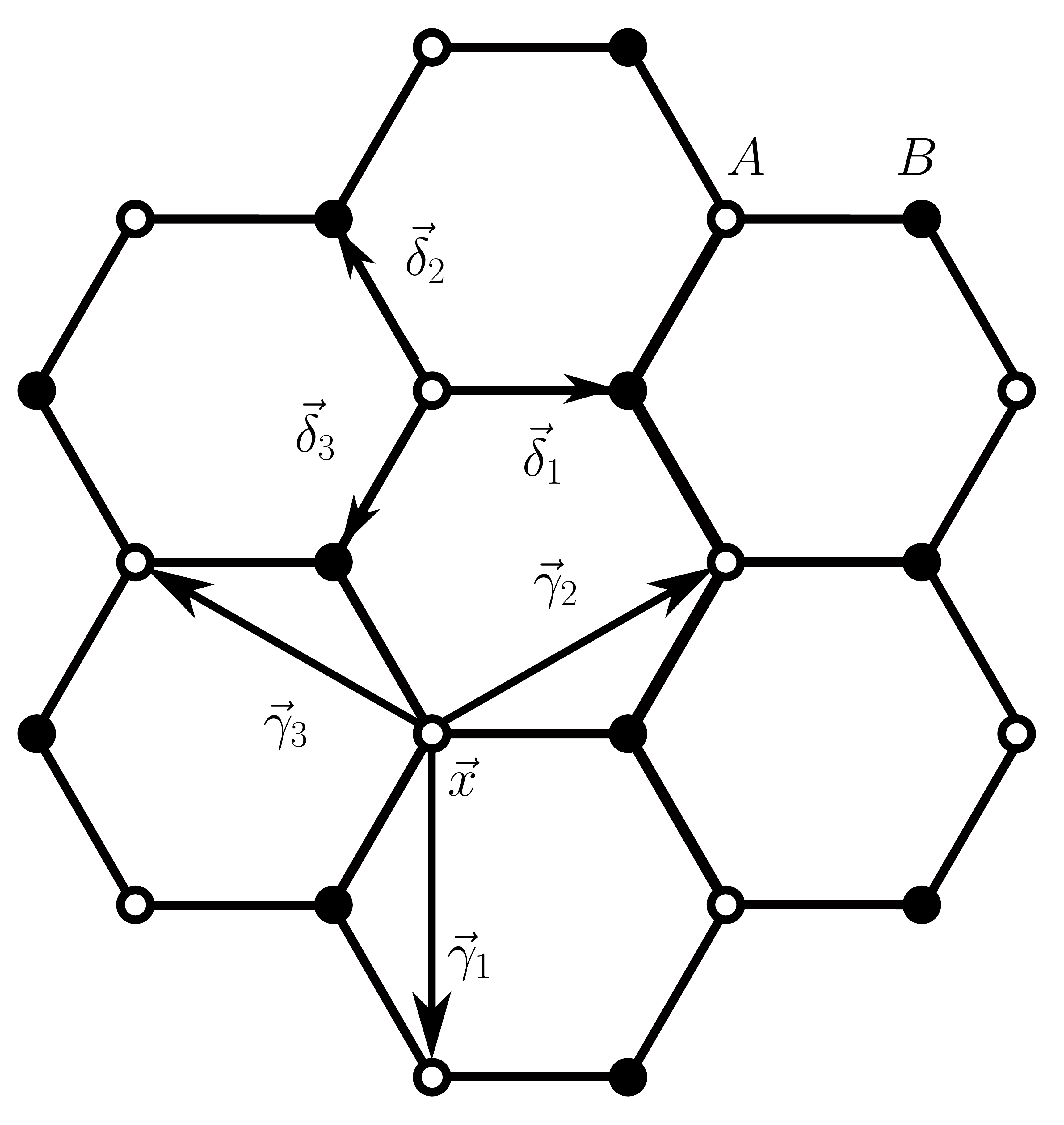}
\caption{The honeycomb lattice of the Haldane-Hubbard model.}
\label{fig3}
\end{figure}

The Haldane-Hubbard model describes interacting fermions on the honeycomb lattice $\L$, which can be understood as the superposition of two triangular sublattices $\L_{A}$ and $\L_{B}$; see Fig. \ref{fig3}. The triangular sublattice $\L_{A}$ is generated by the basis vectors
\be
\vec\ell_{1} = \frac{1}{2}(3, -\sqrt{3})\;,\qquad \vec\ell_{2} = \frac{1}{2}(3, \sqrt{3})\;.
\ee
With each sublattice, we introduce fermionic creation and annihilation operators $a^{\pm}_{\vec x, \s}$, $b^{\pm}_{\vec y,\s}$, where $\s$ is the spin degree of freedom, $\s = \uparrow, \downarrow$. The Hamiltonian is:
\be\label{eq:H}
H = H_{0} + UV - \mu N\;,
\ee
where: $H_{0}$ is the noninteracting Hamiltonian, $UV$ is the Hubbard interaction and $-\mu N$ fixes the chemical potential. The noninteracting Hamiltonian is\cite{H}:
\bea\label{eq:H0}
H_{0} &=& -\sum_{\s=\uparrow\downarrow}\sum_{\langle \vec x, \vec y \rangle} t_{1}\big[ a^{+}_{\vec x, \s} b^{-}_{\vec y,\s} + b^{+}_{\vec y,\s}a^{-}_{\vec x,\s} \big] \nn\\
&& - \sum_{\s=\uparrow\downarrow}\sum_{\langle \langle \vec x, \vec y \rangle \rangle} \big[ t_{2}(\vec x,\vec y) a^{+}_{\vec x,\s}a^{-}_{\vec y,\s} + t_{2}(\vec x,\vec y)^{*} b^{+}_{\vec x,\s}b^{-}_{\vec y,\s} \big]\nn\\
&& + W\sum_{\s=\uparrow\downarrow}\Big[ \sum_{\vec x\in \L_{A}} a^{+}_{\vec x,\s}a^{-}_{\vec x,\s} - \sum_{\vec y\in \L_{B}} b^{+}_{\vec y,\s}b^{-}_{\vec y,\s}\Big]\;;
\eea
the first sum is over nearest-neighbours on $\L$, while the second is over next-to-nearest neighbours. Each site on $\L_{A}$ is connected to its three nearest-neighbours on $\L_{B}$ by the vectors:
\be
\vec \delta_{1} = (1, 0)\;,\quad \vec \delta_{2} = \frac{1}{2}(-1, \sqrt{3})\;,\quad \vec \delta_{3} = \frac{1}{2}(-1, -\sqrt{3})\;.
\ee
The next-to-nearest neighbour hopping parameter $t_{2}(\vec x,\vec y)$ is defined as:
\be
t_{2}(\vec x, \vec x + \vec \gamma_{i}) = e^{i\phi}t_{2}\;,\qquad t_{2}(\vec x, \vec x - \vec \gamma_{i}) = e^{-i\phi}t_{2}\;,
\ee
for $i = 1,2,3$. Explicitely (see Fig. \ref{fig3}):
\be
\vec\gamma_{1} = \vec\ell_{1} - \vec\ell_{2}\;,\qquad \vec\gamma_{2} = \vec \ell_{2}\;,\qquad \vec\gamma_{3} = -\vec\ell_{1}\;.
\ee
The Hubbard interaction term is, as usual:
\be\label{eq:V}
V = \sum_{\vec x} \Big[n_{\vec x,\uparrow}-\frac{1}{2}\Big]\Big[ n_{\vec x,\downarrow} - \frac{1}{2} \Big]\;,
\ee
where the sum ranges over the full honeycomb lattice; the density operator $n_{\vec x,\s}$ is:
\be
n_{\vec x,\s} = \left\{ \begin{array}{cc} a^{+}_{\vec x,\s}a^{-}_{\vec x,\s} & \text{for $\vec x\in \L_{A}$} \\ b^{+}_{\vec x,\s}b^{-}_{\vec x,\s} & \text{for $\vec x\in \L_{B}$}  \end{array} \right.,
\ee
in terms of which we also have $N = \sum_{\vec x,\s} n_{\vec x,\s}$.
The factors $-1/2$ in Eq. (\ref{eq:V}) amount to a redefinition of $\mu$, and simplify the functional integral representation of the model (see Section \ref{sec:funcint}).
%
%

We denote the finite volume version of $H$ by $H_{L}$, with periodic boundary conditions. The finite volume and finite temperature Gibbs state is:
\be
\media{\cdot}_{\b,L}=\Tr\{e^{-\b H_L}\cdot\}/\Tr\{e^{-\b H_L}\}\;,
\ee
and we let
\be
\media{\cdot}_\b=\lim_{L\to\infty}\media{\cdot}_{\b,L}\;,\qquad \media{\cdot} = \lim_{\beta\to\infty} \media{\cdot}_{\b,L}\;.
\ee


\medskip

{\it Correlations, current and conductivity.} It is convenient to define
\be
\Psi^+_{\vec x,\s}= \big(a^+_{\vec x,\s},\, b^+_{\vec x+\vec\d_1,\s}\big)\;,\qquad \Psi^-_{\vec x,\s} = \big(\Psi^+_{\vec x,\s}\big)^\dagger\;;
\ee
also, for any inverse temperature $\b$, we let $\Psi^\pm_{\vec x,\s}(x_0)= e^{H x_0}\Psi^\pm_{\vec x,\s} e^{-H x_0}$ be their evolution
at `imaginary time' $x_0\in[0,\b)$. For general $x_0\in\mathbb R$, we extend $\Psi^\pm_{\vec x,\s}(x_0)$ anti-periodically (of anti-period $\b$) beyond the basic interval $[0,\b)$.
The Fourier transform of the fields is defined as $\Psi^\pm_{\vec x,\s}=\int_\BBB \frac{d\vec k}{|\BBB|} e^{\pm i \vec k\cdot \vec x}
\hat \Psi^\pm_{\vec k,\s}$, where $\BBB$ is the Brillouin zone \cite{brillouin}.
The 2-point function is \bea S_2(\xx,\yy)&=&\media{{\bf T}\,\Psi^-_{\vec x,\s}(x_0)\Psi^+_{\vec y,\s}(y_0)}\nn\\
&=&\int_{\mathbb R\times \BBB}\frac{d\kk}{2\p|\BBB|}e^{-i\kk(\xx-\yy)}\hat S_2(\kk),\nn\eea
where $\xx=(x_0,\vec x)$, $\yy=(y_0,\vec y)$, 
${\bf T}$ is the fermionic time-ordering operator (which orders imaginary times in decreasing order \cite{order}), and $\kk=(k_0,\vec k)$, where $k_0$ is the Matsubara frequency. Note that $S_2$ is a $2\times2$ matrix (with indices in the `sublattice' space) 
and its definition is independent of the 
choice of $\s = \uparrow,\downarrow$.

The current is defined via the Peierls' substitution (see App.\ref{app.ver}), and is equal to 
\be \vec J_{\vec p}(x_0)=\sum_{\s=\uparrow\downarrow}
\int_{\BBB} \frac{d\vec k}{|\BBB|}\hat \Psi^{+}_{\vec k+\vec p,\s}(x_0)\vec M(\vec k,\vec p)\hat \Psi^-_{\vec k,\s}(x_0)\;.\label{eq:curr}\ee
The two components $M_i(\vec k,\vec p)$, $i=1,2$, of $\vec M(\vec k,\vec p)$ are the bare vertex functions, which are $2\times2$ matrices, 
with elements labelled by the spinor indices. For the explicit expression of the bare vertex  functions, see 
App.\ref{app.ver}.

The current-current and the vertex correlations are defined, respectively, as 
\bea && \hat K_{\m\n}(\pp)=\int_{\mathbb R}dx_0 e^{-i p_0x_0}\mmedia{{\bf T}\,J_{\vec p,\m}(x_0); J_{-\vec p,\n}(0)},\nn\\
&& \hat G_{\m}(\kk,\pp)=\int_{\mathbb R}dx_0\int_{\mathbb R}dy_0\, e^{-i p_0x_0+i(k_0+p_0)y_0}\times\nn\\
&&\hskip1.2truecm \times\mmedia{{\bf T}\,J_{\vec p,\m}(x_0);\hat \Psi^-_{\vec k+\vec p,\s}(y_0)
\hat \Psi^+_{\vec k,\s}},\label{eq:kmn}
\eea
where $\m,\n\in\{0,1,2\}$, 
$$J_{\vec p,0}(x_0)=\sum_{\vec x\in\L_A}\sum_{\s=\uparrow\downarrow}e^{-i\vec p\cdot\vec x}\Psi^+_{\vec x,\s}(x_0)M_0(\vec p)\Psi^-_{\vec x,\s}(x_0),$$
with 
\be
M_{0}(\vec p) = \begin{pmatrix} 1 & 0 \\ 0 & e^{-i\vec p_{1}} \end{pmatrix}\;;
\ee
the labels $\mu = 1,2$ refer to the components of the current defined in Eq. (\ref{eq:curr}).
Moreover, $\mmedia{\cdot}=
\lim_{\b\to\infty}\lim_{L\to\infty}L^{-2}\media{\cdot}_{\b,L}$ is the trace per unit volume, and the semi-colon indicates that the expectation is truncated.

For later reference, we also introduce the vertex function:
%
\be\hat \G_{\m}(\kk,\pp)=\hat S_2^{-1}(\kk+\pp)\hat G_{\m}(\kk,\pp)\hat S_2^{-1}(\kk),\label{eq:vf}\ee
where $\hat S_2^{-1}(\kk)$ is the inverse of the 2-point function, thought of as a $2\times2$ matrix. 

Finally, the d.c. Kubo conductivity is defined in terms of the current-current correlation, in units such that $e^{2} = \hbar = 1$, as:
\be
\s_{ij} =- \lim_{p_{0}\rightarrow 0^{+}}\frac{1}{Ap_{0}}\big[\hat K_{ij}(p_0,0)-\hat K_{ij}({\bf 0})\big],
\label{eq:sijdef}\ee
where $i,j=1,2$ and $A=|\vec\ell_1\times\vec\ell_2|=3\sqrt3/2$ is the area of the fundamental cell.

\medskip

{\it Ward Identities.} The continuity equation for the lattice current Eq. \eqref{eq:curr}, when averaged against an arbitrary number of field operators, implies 
exact identities among correlation functions (Ward Identities), valid for any value of the interaction $U$.
In particular, the one relating the 2-point and the vertex functions, which will play an important role in the following, reads as follows: 
\be
\sum_{\m=0}^2 (i)^{\d_{\m,0}}p_\m \hat G_{\m}(\kk,\pp)=\hat S_2(\kk+\pp)M_0(\vec p)-
M_0(\vec p) \hat S_2(\kk).\label{1.14}
\ee
If we derive this equation with respect to $\pp$, compute the result at $\pp={\bf 0}$ and recall the definition \eqref{eq:vf} of the vertex function, we find:
\be \hat \G_{\m}(\kk,{\bf 0})=(-i)^{\d_{\m,0}}\partial_{\m}
\hat S_2^{-1}(\kk)
+\big[\partial_{\m}M_0(\vec 0),\hat S_2^{-1}(\kk)\big].\label{eq:wi-2}\ee
In the following, $\hat \G_{\m}(\kk,{\bf 0})$ will be denoted simply by $\hat \G_{\m}(\kk)$.

\medskip

{\it The non-interacting case.} If $U=0$, the band structure and the phase diagram can be computed explicitly: the Bloch Hamiltonian 
is \cite{H}
\bea &&
\hat H_0(\vec k) =\label{hbloch}\\
&&\quad =
 \begin{pmatrix} -2t_{2}\cos\phi\,\alpha_{1}(\vec k) + m(\vec k) & -t_{1}\Omega^*(\vec k) \\ -t_{1}\Omega(\vec k) &\hskip-.3truecm -2t_{2}\cos\phi\,\alpha_{1}(\vec k) - m(\vec k) \end{pmatrix}
\nonumber\eea
where $\Omega(\vec k) = 1 + e^{-i\vec k\cdot \vec\ell_{1}} + e^{-i\vec k\cdot \vec\ell_{2}}$ and
\bea
&&\alpha_{1}(\vec k) = \sum_{i=1}^3\cos(\vec k\cdot \vec \gamma_{i})\;,\quad \alpha_{2}(\vec k) = \sum_{i=1}^3 \sin(\vec k\cdot \vec \gamma_{i})\nn\\
&&m(\vec k) = W - 2t_{2}\sin\phi\, \alpha_{2}(\vec k)\;.
\eea
%
%
The corresponding energy bands are 
\begin{equation}
\e_{\pm}(\vec k) = -2t_{2}\cos\phi\, \alpha_{1}(\vec k) \pm \sqrt{m(\vec k)^{2} + t_{1}^{2}|\Omega(\vec k)|^{2}}\;.\nn
\end{equation}
To make sure that the energy bands do not overlap, we assume that $t_2/t_1<1/3$.
The two bands can only touch  at the {\it Fermi points} $\vec p_{F}^{\pm} = \big( \frac{2\pi}{3}, \pm \frac{2\pi}{3\sqrt{3}} \big)$, which are the two zeros of $\Omega(\vec k)$,
around which $\O(\vec p_F^\pm+\vec k')\simeq \frac32(i k_1'\pm k_2')$.
The condition that the two bands touch
at $\vec p_F^\o$, with $\o = +,-$, is that $m_\o=0$, with 
\be
m_{\o} \equiv m(\vec p_{F}^{\o}) = W +\o 3\sqrt{3}\,t_{2}\sin\phi\;.\nn
\ee
Therefore, the unperturbed critical curves are given by the values of $(\phi, W)$ such that:
\be
W=\pm 3\sqrt3\, t_2\sin \phi\,,
\ee
which correspond to the dotted curves in Fig.\ref{figaltezza}. Fixing the chemical potential in such a way that the Fermi energy lies in between the two bands, 
\be
\m=-2t_2\cos\phi\,\a_1(\vec p_F^\pm)=-3 t_2\cos\phi\;,
\ee
 the system passes from a semi-metallic behavior, when $(\phi,W)$ is on the critical line, to an insulating behavior, characterized by the exponential decay of correlations, 
when $W\not=\pm3\sqrt3 t_2\,\sin\phi$.

The insulating phase consists of four disconnected regions in the $(\phi,W)$ plane, two of which are `topologically trivial', while the other two have non-zero Hall conductivity, see Fig.\ref{figaltezza}: more precisely, 
if $W\not=\pm3\sqrt3 t_2\,\sin\phi$,
\be \sigma_{12} = \frac{\nu}{2\pi}\;,\quad \nu = \text{sign}(m_{-}) - \text{sign}(m_{+}) \;.\nn
\ee

\section{Renormalization Group analysis}\label{sec3}

We now construct the interacting correlations and phase diagram, by using a convergent renormalized expansion, in the spirit of Ref.\cite{GM,GMP,GMPhall}. In this section, we introduce the 
functional integral formulation of the model, discuss the exact lattice symmetries of the fermionic action, and describe the infrared integration, including the study of the flow of the running coupling constants. 
One of the main results of this section is the equation for the interacting critical line.

\subsection{Functional integral formulation}\label{sec:funcint}

We are interested in the semi-metallic and insulating regimes of the interacting system. We, therefore, set the chemical potential accordingly (its value will be different, in general, from the unperturbed one):
\be
\m=-2 
t_2\cos\phi\,\a_1(\vec p_F^\pm)-\xi,\nn\ee 
where $\xi$ (the shift of the chemical potential) must be chosen as a function of $U,W,\phi,$ so that the renormalized propagator either 
has a linear, `conical', infrared singularity (along the interacting critical line),
or is gapped (in the insulating phase). 

The generating function $\WW(f,A)$ for correlations, in which $f$ is the external field conjugated to the fermionic fields, and $A$ is the external field conjugated to 
the current, can be written as the following Grassmann integral:
\be
e^{\WW(f,A)}=\frac{\int P(d\psi)e^{-V(\psi)+(\psi,f)+(j,A)}}{\int P(d\psi)e^{-V(\psi)}},\label{eq.11}
\ee
where: $\psi^\pm_{\xx,\s}$, with $\xx=(x_0,\vec x)\in\mathbb R\times \L_A$ and $\s\in\{\uparrow,\downarrow\}$, 
is a two-component Grassmann spinor (it is the Grassmann counterpart of $\Psi^\pm_{\vec x,\s}(x_0)$),
whose components will be denoted by $\psi^\pm_{\xx,\s,\r}$, with $\r=1,2$;
$P(d\psi)$ is the fermionic Gaussian integration with propagator
\be g(\xx,\yy)=\int_{\mathbb R\times\BBB}\frac{d\kk}{2\p|\BBB|}
e^{-i\kk(\xx-\yy)} \hat g(\kk),\label{prop}\ee
where, letting $R(\vec k)= -2t_2\cos\phi\big(\a_1(\vec k)-\a_1(\vec p_F^{\pm})\big)$, 
\be \hat g(\kk)=\begin{pmatrix} 
&-ik_0+R(\vec k)+ m(\vec k)  & -t_1 \O^*(\vec k)\\  &-t_1 \O(\vec k)
 & -i k_0+R(\vec k)- m(\vec k)
\end{pmatrix}^{\!\!\!-1}\nn
\ee
and, at contact, $g(\xx,\xx)$ should be interpreted as $\lim_{\e\to 0^+}[g(\xx+(\e,\vec 0),\xx)+g(\xx-(\e,\vec 0),\xx)]$;
\be
V(\psi)=\int_{\mathbb R} dx_0\sum_{\vec x\in\L_A} \sum_{\rho=1,2}\big(Un^\r_{\xx,\uparrow}n^\r_{\xx,\downarrow}+\xi\sum_{\s=\uparrow,\downarrow} n^\r_{\xx,\s}\big),\nn
\ee
where $n^\r_{\xx,\s}=\psi^+_{\xx,\s,\r}\psi^-_{\xx,\s,\r}$;
and, finally, 
\bea && (\psi,f)=\int_{\mathbb R}dx_0\sum_{\vec x\in\L_A}\sum_{\s=\uparrow\downarrow}(\psi^+_{\xx,\s}f^-_{\xx,\s}+f^+_{\xx,\s}\psi^-_{\xx,\s}),\nn\\
&& (j,A)=\int_{\mathbb R^3} \frac{d\pp}{(2\p)^3}\hat A_{\pp,\m}\hat\jmath_{\pp,\m},\nn\eea
where $\hat\jmath_{\pp,\m}=\sum_{\s=\uparrow\downarrow}\int_{\mathbb R\times\BBB} \frac{d\kk}{2\p|\BBB|}\hat\psi^+_{\kk+\pp,\s}\G_\m(\vec k,\vec p)\hat \psi^-_{\kk,\s}$,
in which $\G_\m(\vec k,\vec p)$ are the {\it bare} vertex functions, namely: $\G_0(\vec k,\vec p)=M(\vec p)$, and, if $i=1,2$, $\G_i(\vec k,\vec p)$ are the two components of the (matrix-valued) vector $\vec M(\vec k,\vec p)$ defined in \eqref{eq:curr} and following lines.  
In terms of these definitions, the correlations can be re-expressed as
\bea
S_2(\xx,\yy) &=& \frac{\partial^2\mathcal W}{\partial f^+_{\xx,\s}\partial f^-_{\yy,\s}}(0,0)\;,\nn\\
K_{\m\n}(\xx,\yy) &=& \frac{\partial^2\mathcal W}{\partial j_{\xx,\m}\partial j_{\yy,\n}}(0,0)\;,
\eea
and of suitable linear combinations of 
\be
G_{2,1;\m}(\xx,\yy,\zz)=\frac{\partial^3\mathcal W}{\partial A_{\xx,\m}\partial f^+_{\yy,\s}\partial f^-_{\zz,\s}}(0,0)\;.
\ee

We now compute the generating function Eq. \eqref{eq.11} via a renormalized expansion, which is convergent uniformly close to (and even on) the critical line. Note that, on this line, the Grassmann integral has an infrared problem.
In order to resolve and re-sum the corresponding singularities, we proceed in a multi-scale fashion. 
First of all, we distinguish the ultraviolet modes, corresponding to large values of the Matsubara frequency, from the infrared ones, by introducing two 
compactly supported cut-off functions, $\chi_\pm(\kk)$, supported in the vicinity of the Fermi points (more precisely, we let $\c_\pm(\kk)=\c_0(\kk-\pp_F^\pm)$, where
$\c_0$ is a smoothed out characteristic function of the ball of radius $a_0$, with $a_0$ equal to, say, $1/3$,
and $\pp_F^\pm=(0,\vec p_F^\pm)$)
and by letting $\chi_{uv}(\kk)=1-\sum_{\o=\pm}\chi_\o(\kk)$.
We correspondingly split the propagator in its ultraviolet and infrared components: 
\be
g(\xx,\yy)=g^{(1)}(\xx,\yy)+\sum_{\o=\pm} e^{-i\vec p_F^\o(\vec x-\vec y)}g_{\o}^{(\le 0)}(\xx,\yy)\label{eq:4}
\ee
where $g^{(1)}(\xx,\yy)$ and $g_{\o}^{(\le 0)}(\xx,\yy)$ are defined in a similar way as Eq. \eqref{prop}, with $\hat g(\kk)$ replaced by $\chi_{uv}(\kk) \hat g(\kk)$ and by $\chi_{0}(\kk) \hat g(\kk+\pp_F^\o)$,
respectively.  We then split the Grassmann field as a sum of two independent fields, with propagators $g^{(1)}$ and $g^{(\le 0)}$: 
$$\psi_{\xx,\s}^\pm=\psi^{\pm(1)}_{\xx,\s}+\sum_{\o=\pm }e^{\pm i\vec p_F^\o\vec x} \psi_{\xx,\s,\o}^{\pm(\le0)}$$
and we rewrite the Grassmann Gaussian integration as the product of two independent Gaussians: $P(d\psi)=P(d\psi^{(\le 0)})P(\psi^{(1)})$. By construction, the integration of the `ultraviolet' field $\psi^{(1)}$ does not have any infrared singularity and, therefore, can be performed in a straightforward manner, thus allowing us to rewrite the generating function $\mathcal W(f,A)$ as the logarithm of 
\be \frac{e^{\WW^{(0)}(f,A)}}{\mathcal N_0}\int P(d\psi^{(\le 0)})e^{-V^{(0)}(\psi^{(\le 0)})+B^{(0)}(\psi^{(\le 0)}, f, A)},\label{eq:v0}
\ee
where $V^{(0)}$ and $B^{(0)}$ are, respectively, the effective potential and the effective source (which depend explicitly on, respectively, $\psi^{(\le 0)}$ and $\psi^{(\le 0)},f,A$), 
$\WW^{(0)}$ is independent of $\psi^{(\le 0)}$ (and depends explicitly on $f,A$), and $\mathcal N_0=\int P(d\psi^{(\le 0)})e^{-V^{(0)}(\psi^{(\le 0)})}$.
Both $V^{(0)}$ and $B^{(0)}$ are expressed as
series of monomials in the $\psi,f,A$ fields, whose kernels (given by the sum of all possible Feynman diagrams with fixed number and fixed location of the external legs)
are {\it analytic functions} of the interaction strength, for $U$ sufficiently small. The proof of their analyticity is based on a determinant expansion and on a systematic use of the Gram-Hadamard bounds, see 
Ref.\cite{GM,GMPhall}.

\subsection{Symmetries}\label{sec.symm}

Before tackling the multi-scale integration of the infrared modes, we make a digression about the symmetry structure of the effective potential, and in particular of its local parts: 
the purpose is to classify the possible relevant and marginal coupling constants.
In the case $t_2=W=\mu=0$ (standard graphene model) the lattice symmetries severely constrain the form of the quadratic terms in the effective potential: in particular, the interaction does not 
shift the chemical potential, nor does it generate a mass\cite{GM, GMP,FA,HJR}. In the general case ($W,t_2,\phi\neq0$) the model is invariant under the following symmetry 
transformations (since they do not mix the spin indices, for notational convenience 
we temporarily drop the spin labels from the formulas). 

We discuss the symmetries in the {\it absence} of external fields, since we will use them only to infer the structure of the 
relevant and marginal contributions to the effective potential $V^{(0)}$. Once the structure of these terms is known, the structure of the marginal contributions to the effective source $B^{(0)}$ can be computed by using the Ward Identity \eqref{eq:wi-2}.

\medskip

(1) {\it Discrete rotations:}
\be
\hat \psi_{\kk}^-\to e^{i\vec k(\vec \d_3-\vec \d_1)n_-}\hat \psi_{T\kk}^-\;,\quad 
\hat\psi_{\kk}^+\to\hat\psi_{T\kk}^+e^{-i\vec k(\vec \d_3-\vec \d_1)n_-}
\ee
where, denoting the Pauli matrices by $\s_1,\s_2,\s_3$, we defined
\be
n_-=(1-\s_3)/2\;,\qquad T\kk=(k_0,e^{-i\frac{2\p}{3}\s_2}\vec k)\;;
\ee
that is, $T$ is the spatial rotation by $2\p/3$ in the counter-clockwise direction.

\medskip

(2) {\it Complex conjugation}:
\be
\hat\psi^{\pm}_{\kk}\rightarrow \hat\psi^{\pm}_{-\kk}\;,
\ee
combined with 
\be
c\rightarrow c^{*}\;,\quad \phi\to-\phi\;,
\ee
where $c$ is a generic constant appearing in $P(d\psi)$ or in $V(\psi)$.

\medskip

(3) {\it Horizontal reflections}:
\be
\hat\psi^{-}_{\kk}\to \s_1\hat\psi^-_{R_h\kk}\;,\quad \hat\psi^{+}_{\kk}\to \hat\psi^+_{R_h\kk}\s_1\;,
\ee
with 
\be
R_h\kk=(k_0,-k_1,k_2)\,,\quad (W,\phi)\to (-W,-\phi)
\ee

\medskip

(4) {\it Vertical reflections}:
\be
\hat\psi^{\pm}_{\kk}\rightarrow
\hat\psi^{\pm}_{R_v\kk}\,,
\ee
with 
\be
R_v\kk=(k_0,k_1,-k_2)\;,\quad \phi\to-\phi.
\ee

(5) {\it Particle-hole}:
\be
\hat\psi^{-}_{\kk}\to i\hat\psi^{+,T}_{P\kk}\;,\quad \hat\psi^{+}_{\kk}\to i\hat\psi^{-,T}_{P\kk}\;,
\ee
with
\be
P\kk=(k_0,-k_1,-k_2)\;,\quad \phi\to-\phi\;.
\ee

Note that, at fixed $W,\phi$, the theory is invariant under the transformations (1), (2)+(4), and (2)+(5). In particular, these transformations leave the quadratic part 
$Q^{(0)}(\psi)=\sum_\s\int \frac{d\kk}{(2\p|\BBB|)}\hat \psi^+_{\kk,\s}\hat W_2(\kk)\hat \psi^-_{\kk,\s}$
of the effective potential $V^{(0)}(\psi)$ invariant. This means that:
\bea \hat W_2(\kk)&=&e^{-i\vec k(\vec \d_1-\vec\d_2)n_-} \hat W_{2}(T^{-1}\kk)e^{i\vec k(\vec \d_1-\vec\d_2)n_-}\nonumber\\
&=& \hat W_{2}^{*}(-k_0,-k_1,k_2)\label{symm}\\
&=&\hat W_{2}^{\dagger}(-k_0,k_1,k_2).\nonumber\eea
As we will see in the next section, the values of $\hat W_2(\kk)$ and of its derivatives at the Fermi points define the {\it effective coupling constants}. By \eqref{symm}, we find, for $\o=\pm$,
\bea \hskip-.2truecm
\hat W_2(\pp_F^\o)&=&e^{-i\frac{2\p}3\o n_-}\hat W_2(\pp_F^\o)e^{i\frac{2\p}3\o n_-}\nonumber\\
&=& \hat W_{2}^{*}(\pp_F^\o)=
\hat W_{2}^{\dagger}(\pp_F^\o),\nonumber\eea
which implies that \be \hat W_2(\pp_F^\o)=\xi_{\o}+\d_{\o}\s_3,\label{eq:w2_loc0}\ee
for two {\it real} constants $\xi_{\o}$ and $\d_{\o}$.

If we derive \eqref{symm} with respect to $\kk$ and compute the result at $\pp_F^\o$, we find:
\bea 
\partial_\kk\hat W_2(\pp_F^\o)&=&e^{-i\frac{2\p}3\o n_-}T\partial_\kk\hat W_2(\pp_F^\o)e^{i\frac{2\p}3\o n_-}\nonumber\\
&=&(-R_v)\partial_\kk \hat W_{2}^{*}(\pp_F^\o)\label{symm_der}\\
&=&(-P)\partial_\kk
\hat W_{2}^{\dagger}(\pp_F^\o),\nonumber\eea
where $R_v$ (resp. $P$) is the diagonal matrix with diagonal elements $(1,1,-1)$ (resp. $(1,-1,-1)$).
By using \eqref{symm_der}, it is straightforward to check that
\be \kk'\partial_{\kk}\hat W(\pp_{F}^{\o}) =
\begin{pmatrix}  -i z_{1,\o} k_{0} & -u_{\o}(-i k_{1}' +\o k_{2}') \\ -u_{\o}(i k_{1}' + \o k_{2}') & -i z_{2,\o} k_{0}  \end{pmatrix},\label{eq:6}\ee
where $\kk'=\kk-\pp_F^\o=(k_0,\vec{k'})$, and $u_\o, z_{1,\o},z_{2,\o}$ are {\it real} constants. In conclusion, for general values of $W,\phi$, the linearization of $\hat W_2(\kk)$ at $\pp_F^\o$ is parametrized by 5 real constants,  namely 
$\xi_{\o}, \d_{\o}, u_\o, z_{1,\o}$ and $z_{2,\o}$, the first two are relevant coupling constants, and the other three are marginal. Note that, in general,
the values of these constants depend on $\o$ (therefore, there are 5 of them at $\pp_F^+$ and 5 more at $\pp_F^-$). Note also that, in general, $z_{1,\o}\neq z_{2,\o}$, i.e., the wave function renormalization 
depends explicitly on the spinor index, an effect that can be checked explicitly at second order in perturbation theory (see below), and cannot be explained purely in terms of the 
relativistic approximation of the model around the Fermi points. 

Note that there are special points in the $(W,\phi)$ plane, for which the model has more symmetries, and where the number of independent couplings is smaller than in the general case. For instance, if $W=\phi=0$, the model is invariant under 
all the 5 symmetry transformations listed above, in which case it is straightforward to see that 
\bea
&&\xi_\o=\xi_{-\o}\;,\quad \d_\o=0\;,\quad u_\o=u_{-\o}\;,\nn\\
&&z_{1,\o}=z_{2,\o}=z_{1,-\o}=z_{2,-\o}\;.
\eea
 A similar discussion applies to the case $W=0$, $\phi=\p$.

Finally, if $\phi=\p/2$, the model is invariant under the following additional symmetry transformation (see also Ref.\cite{Z}): 
\be
\hat\psi^{-}_{\kk,\s}\to-i \s_1\s_3\hat\psi^-_{-R_v\kk,\s}\;,\quad \hat\psi^{+}_{\kk,\s}\to -i\hat\psi^+_{-R_v\kk,\s}\s_3\s_1\;,
\ee
which implies that 
$$\hat W_2(\kk)=-\s_3\s_1\hat W_2(-k_0,-k_1,k_2)\s_1\s_3,$$
so that, in particular, 
\be
\xi_\o=0\;,\quad z_{1,\o}=z_{2,\o}\;.
\ee
A similar discussion applies to $\phi=-\pi/2$.

\subsection{Infrared integration}\label{sec.RG}

Let us now describe the integration of the infrared fields. We shall focus on the semi-metallic behavior of the system at, or very close to, a generic point of the critical line. 
Moreover, since we are interested in the behavior of the current-current correlations around $\pp={\bf 0}$, we shall assume that the external field $\hat A_{\pp,\m}$ is supported in the vicinity of the origin (in 
particular, we assume that it vanishes in the vicinity of $\pp_F^{\o}-\pp_F^{-\o}$, $\o=\pm$). 

By dimensional considerations, the quadratic terms in the effective action are {\it relevant}, and, the ones corresponding to the renormalization of the mass are of particular importance. The flow 
of the effective mass tends to diverge linearly under the RG iterations, which signals that, in general, the location of the critical lines is changed by the interaction. In order to construct a convergent expansion, 
we need to dress the mass, after which we determine the location of the renormalized critical lines, which is given by the condition that the dressed mass vanishes. 

More in detail, we proceed as follows. We perform the integration of the infrared modes in \eqref{eq:v0} iteratively, by 
decomposing the fermionic fields as $\psi_{\xx,\s,\o}^{\pm(\le0)}$ as $\psi_{\xx,\s,\o}^{\pm(\le0)}=\sum_{h\le 0}  \psi_{\xx,\s,\o}^{\pm(h)}$, where $\psi_{\xx,\s,\o}^{\pm(h)}$ is a Grassmann 
field whose propagator is supported on the momenta $\kk$ such that $|\kk-\pp_F^\o|\sim 2^h$, and by integrating the fields $\psi^{(0)},\psi^{(-1)},\ldots$ step by step.
After the integration of the modes on scales $0,-1,\ldots, h+1$, we rewrite the generating function $\WW(f,A)$ as the logarithm of 
\be \frac{e^{\WW^{(h)}(f,A)}}{\mathcal N_h}\int P(d\psi^{(\le h)})e^{-V^{(h)}(\psi^{(\le h)})+B^{(h)}(\psi^{(\le h)}, f, A)},\label{eq:vh}
\ee
where $V^{(h)}$ and $B^{(h)}$ are, respectively, the effective potential and source terms, to be defined inductively in the following. Moreover, 
$P(d\psi^{(\le h)})$ is the Grassmann Gaussian integration with propagator (diagonal with respect to the $\s$ and $\o$ indices)
\bea g^{(\le h)}_\o(\xx,\yy)&=&\int P(d\psi^{(\le h)})\psi^{-(\le h)}_{\xx,\s,\o}\psi^{+(\le h)}_{\yy,\s,\o}\nn\\
&=&
\int\frac{d\kk'}{2\p|\BBB|}
e^{-i\kk'(\xx-\yy)} \hat g_\o^{(\le h)}(\kk'),\nn\eea 
where $\kk'=(k_0,\vec k')$ and, letting $r_\o(\vec k')= R(\vec k'+\vec p_F^\o)$, 
$s_\o(\vec k')=-t_1[\O(\vec k'+\vec p_F^{\,\o})-\frac32(ik_1'+\o k_2')]$ and $\c_h(\kk')=\c_0(2^{-h}\kk')$ (here $\c_0$ is the cutoff function defined 
a few lines before \eqref{eq:4}),
\be\hat g_\o^{(\le h)}(\kk)=\c_h(\kk') \begin{pmatrix} a_{1,\o,h}(\kk') & b^*_{\o,h}(\kk')\\
b_{\o,h}(\kk') & a_{2,\o,h}(\kk')\end{pmatrix}^{\!\!\!-1},\label{eq:prop.2}\ee
with
\bea && a_{\r,\o,h}(\kk)=-ik_0Z_{\r,\o,h}+r_\o(\vec k')+(-1)^{\r-1} m_{\o,h}(\vec k'),\nn\\
&& b_{\o,h}(\kk')=-v_{\o,h} (ik_1'+\o k_2')+s_\o(\vec k')\;.\label{eq:prop.3}
\eea
in which $Z_{j,\o,h}$, $m_{\o,h}(\vec k')$ and $v_{\o,h}$ are, respectively, the wave function renormalizations, the effective mass and effective velocities, to be defined inductively in the following.
Their initial values are: 
\be
Z_{j,\o,0}=1\;,\quad m_{\o,0}(\vec k')=m(\vec k'+\vec p_F^{\,\o})\;,\quad v_{\o,0}=\frac32t_1.
\ee

In order to clarify the inductive definition of the effective potential, source, etc, we now describe the integration step at scale $h$. We start from \eqref{eq:vh}, where 
$V^{(h)}(\psi)$ is a sum of even monomials in the $\psi$ fields, whose kernels of order $n$ are denoted by 
$W^{(h)}_{n}$ (for notational simplicity, we temporarily drop the 
space-time, spin, spinor and valley indices of the fermionic fields). Similarly, we denote the kernels of $B^{(h)}$ of order $n$ in $\psi$, $m$ in $f$ and $q$ in $A$, by $W_{n,m,q}$.
The scaling dimension of the kernels $W_{n}$ and $W_{n,m,q}$
is (see Ref.\cite{GM,GMP,GMPhall}) 
\be D=3-n-m-q,\label{eq:scdim}\ee
with the convention that $D>0$ corresponds to relevant, $D=0$ to marginal, and $D<0$ to irrelevant operators.
Note that the only relevant terms are those with $n+m=2$, and the only marginal terms are those with $n+m=2$ and $q=1$ (note that, by construction, $n+m$ is positive and even).
In particular, the effective electron-electron interaction, corresponding to the case $n=4$ and $m=q=0$, is {\it irrelevant}.

In order to define a convergent renormalized expansion, we need to re-sum the relevant and marginal terms. For this purpose, we split $V^{(h)}$ and $B^{(h)}$ into their {\it local} and {\it irrelevant} parts (here, for 
simplicity, we spell out the definitions only in the $f=0$ case, the general case is treatable analogously, along the lines of, e.g., Sect. 12 of Ref.\cite{GeM}, or Ref.\cite{GMP}): 
$V^{(h)}=\LL V^{(h)}+\RR V^{(h)}$ and $B^{(h)}=\LL B^{(h)}+\RR B^{(h)}$, where, denoting the quadratic part of $V^{(h)}$ by $$\sum_{\o,\s}\int \frac {d\kk'}{2\p|\BBB|}\hat \psi^+_{\kk',\s,\o} \hat W^{(h)}_{2;\o}(\kk')
\hat \psi^-_{\kk',\s,\o},$$ and
the part of $B^{(h)}$ of order $(2,0,1)$ in $(\psi,f,A)$ by $$\sum_{\o,\s}\int\frac{d\pp}{(2\p)^3}\int \frac {d\kk'}{2\p|\BBB|}\hat A_{-\pp,\m}\hat \psi^+_{\kk'+\pp,\s,\o} \hat W^{(h)}_{2,1;\m,\o}(\kk',\pp)\hat \psi^-_{\kk',\s,\o}$$
we let:
\bea && \LL V^{(h)}(\psi)=\sum_{\o=\pm}\sum_{\s=\uparrow\downarrow}\int \frac {d\kk'}{2\p|\BBB|}\times\nn\\
&&\quad \times\hat \psi^+_{\kk',\s,\o} [\hat W^{(h)}_{2;\o}({\bf 0})+\kk'\partial_{\kk'}\hat W^{(h)}_{2;\o}({\bf 0})\big]\hat \psi^-_{\kk',\s,\o},\nn\eea
and 
\bea
&& \LL B^{(h)}(\psi,0,A)=\sum_{\o=\pm}\sum_{\s=\uparrow\downarrow}\sum_{\m=0}^2\int\frac{d\pp}{(2\p)^3}\int \frac {d\kk'}{2\p|\BBB|}\times\nn\\
&&\quad \times\hat A_{\pp,\m}\hat \psi^+_{\kk'+\pp,\s,\o} \hat W^{(h)}_{2,1;\m,\o}({\bf 0},{\bf 0})\hat \psi^-_{\kk',\s,\o}.\nn
\eea

By the symmetries discussed in the previous section (see, in particular, \eqref{eq:w2_loc0} and \eqref{eq:6})
\bea && \LL V^{(h)}(\psi)=\sum_{\o=\pm}\int \frac {d\kk'}{2\p|\BBB|}\Big[2^h\xi_{\o,h}\hat \psi^+_{\kk',\s,\o} \hat \psi^-_{\kk',\s,\o}+\label{eq:lv}\\
&&+\hat \psi^+_{\kk',\s,\o} \begin{pmatrix} -i z_{1,\o,h} k_{0}+\d_{\o,h} & -u_{\o,h}(-i k_{1}' +\o k_{2}') \\ -u_{\o,h}(i k_{1}' + \o k_{2}') & -i z_{2,\o,h} k_{0}-\d_{\o,h}  
\end{pmatrix}
\hat \psi^-_{\kk',\s,\o}\Big],\nn\eea
where $\xi_{\o,h}, \d_{\o,h},z_{j,\o,h}, u_{\o,h}$ are real constants.
Moreover, by using the Ward Identity \eqref{eq:wi-2}, we find that 
\bea && \LL B^{(h)}(\psi,0,A)=\sum_{\o=\pm}\sum_{\s=\uparrow\downarrow}\sum_{\m=0}^2\int\frac{d\pp}{(2\p)^3}\int \frac {d\kk'}{2\p|\BBB|}\times\nn\\
&&\quad \times \hat A_{\pp,\m}\hat \psi^+_{\kk'+\pp,\s,\o}
\g_{\m,\o,h}\hat \psi^-_{\kk',\s,\o},\label{eq:lbh}\eea
where 
\bea
\g_{0,\o,h}&=&-\sum_{\r=1}^2(Z_{\r,\o,h}+z_{\r,\o,h})n_\r\nn\\
\g_{1,\o,h}&=&-(v_{\o,h}+u_{\o,h})\s_2\nn\\
\g_{2,\o,h}&=&-\o(v_{\o,h}+u_{\o,h})\s_1
\eea 
in which $n_\r=(1+(-1)^{\r-1}\s_3)/2$ and $\s_i$ are the standard Pauli matrices. 

Once the effective potential and source have been split into local and irrelevant parts, we combine the part of $\LL V^{(h)}$ in the second line of \eqref{eq:lv} with the 
Gaussian integration $P(d\psi^{(\le h)})$, thus defining a dressed measure $\tilde P(d\psi^{(\le h)})$ whose propagator $\tilde g^{(\le h)}_\o(\xx,\yy)$ is 
analogous to $g^{(\le h)}_\o(\xx,\yy)$, with the only difference that the functions $a_{\r,\o,h}$, $b_{\o,h}$ in \eqref{eq:prop.2}-\eqref{eq:prop.3}
are replaced by \bea \tilde a_{\r,\o,h-1}(\kk)&=&-ik_0\tilde Z_{\r,\o,h-1}(\kk')+r_\o(\vec k')\nn\\
&&+(-1)^{\r-1} \tilde m_{\o,h-1}(\kk'),\nn\\
\tilde b_{\o,h-1}(\kk')&=&-\tilde v_{\o,h-1}(\kk') (ik_1'+\o k_2')+s_\o(\vec k'),\nn\eea 
with \bea &&\tilde Z_{\r,\o,h-1}(\kk')=Z_{\r,\o,h}+z_{\r,\o,h}\,\c_h(\kk'),\nn\\
&&\tilde m_{\o,h-1}(\kk')=m_{\o,h}(\vec k')+\d_{\o,h}\,\c_h(\kk'),\nn\\
&& \tilde v_{\o,h-1}(\kk')=v_{\o,h}+u_{\o,h}\,\c_h(\kk').\nn\eea
Now, by rewriting the support function $\c_h(\kk')$ in the definition of $\tilde g^{(\le h)}_\o(\xx,\yy)$ as $\c_h(\kk')=f_h(\kk')+\c_{h-1}(\kk')$,
we correspondingly rewrite: $\tilde g^{(\le h)}_\o(\xx,\yy)=\tilde g^{(h)}_\o(\xx,\yy)+g^{(\le h-1)}_\o(\xx,\yy)$, where $g^{(\le h-1)}_\o(\xx,\yy)$ is defined exactly as in 
\eqref{eq:prop.2}-\eqref{eq:prop.3}, with $h$ replaced by $h-1$, and $Z_{\r,\o,h-1}, m_{\o,h-1}, v_{\o,h-1}$ defined by the flow equations:
\bea &&Z_{\r,\o,h-1}=Z_{\r,\o,h}+z_{\r,\o,h},\nn\\
&& m_{\o,h-1}(\vec k')=m_{\o,h}(\vec k')+\d_{\o,h},\label{eq:flow.1}\\
&& v_{\o,h-1}=v_{\o,h}+u_{\o,h}.\nn\eea
At this point, we integrate the fields on scale $h$, and define:
\bea &&e^{-V^{(h-1)}(\psi)+B^{(h-1)}(\psi,f,A)+w^{(h)}(f,A)}=C_h\int \tilde P(d\psi^{(h)})\times\nn\\
&&\quad \times e^{-F_\xi^{(h)}(\psi^{(h)}+\psi)+\RR V^{(h)}(\psi^{(h)}+\psi)+B^{(h)}(\psi^{(h)}+\psi, f, A)},\nn\eea
where $\tilde P(d\psi^{(h)})$ is the Gaussian integration with propagator $\tilde g^{(h)}_\o$, $F_\xi^{(h)}(\psi)=
\sum_{\o}2^h\xi_{\o,h}\int \frac {d\kk'}{2\p|\BBB|}\hat \psi^+_{\kk',\s,\o} \hat \psi^-_{\kk',\s,\o}$, and $C_h^{-1}=
\int \tilde P(d\psi^{(h)})e^{-F_\xi^{(h)}(\psi^{(h)})+\RR V^{(h)}(\psi^{(h)})}$. Finally, letting $\mathcal W^{(h-1)}=\mathcal W^{(h)}+w^{(h)}$, we obtain the 
same expression as \eqref{eq:vh}, with $h$ replaced by $h-1$. This concludes the proof of the inductive step, corresponding to the integration of the fields on scale $h$.

The integration procedure goes on like this, as long as the two effective masses $m_{\pm,h}$ are small, as compared to $2^h$. If we are not exactly at the `graphene point' $W=\phi=0$, i.e., 
if we are close to, or at, any other point on the critical line but the origin, then after a while we reach a scale $h_1$ at which $\max_\o|m_{\o,h_1}|\equiv |m_{\o_1,h_1}|\simeq 2^{h_1}$
(possibly, $h_1=0$, in the case that $\max_\o|m_\o|$ is of order 1, i.e., if $W,\phi$ are far enough from the graphene point). Note that, once we reach scale $h_1$, the field $\psi_{\kk',\s,\o_1}^{(\le h_1)}$ is 
massive `on the right scale' $2^{h_1}$.
At that point, we integrate out the field $\psi_{\kk',\s,\o_1}^{(\le h_1)}$ in a single step, and we are left with a (chiral) theory, whose only dynamical degree of freedom is $\psi^{(\le h_1)}_{\kk',\s,\o_2}$,
with $\o_2=-\o_1$.

From that scale on, we integrate $\psi^{(\le h_1)}_{\kk',\s,\o_2}=\sum_{h\le h_1}\psi^{(h)}_{\kk',\s,\o_2}$ in a multi-scale fashion, analogous to the one discussed above, with the important difference that 
only the running coupling constants corresponding to the valley index $\o=\o_2$ continue to flow. The multi-scale integration goes on until we reach a scale $h_2$ such that $|m_{\o_2,h_2}|\simeq 2^{h_2}$,
at which point we can integrate out the remaining degrees of freedom in a single step. The criticality condition, i.e., the condition that the system is on the (renormalized) critical line, corresponds to 
the condition that $h_2=-\infty$.

\subsection{The flow of the running coupling constants}

The multi-scale integration described in the previous section defines a flow for the effective chemical potential $\n_{\o,h}$, the effective mass $m_{\o,h}=m_{\o,h}(\vec 0)$, the effective 
wave function renormalization $Z_{\r,\o,h}$, and the effective Fermi velocity $v_{\o,h}$. The flow of $m_{\o,h}, Z_{\r,\o,h}$ and $v_{\o,h}$ is driven by Eqs.(\ref{eq:flow.1}), while 
$$\xi_{\o,h-1}=2 \xi_{\o,h}+\b^\xi_{\o,h},$$
where $\b^h_\xi$ is the ($\xi$-component of the) beta function, which is defined in terms of the sum of all the local quadratic contributions in renormalized perturbation theory, and should be thought of as a 
function 
of $U$ and of the sequence of the effective coupling constants. Remember that the flow drives the effective couplings with both $\o=+$ and $\o=-$, up to the scale $h_1$; then the flow 
of the couplings with $\o=\o_1$ is stopped, and only the couplings with $\o=\o_2$ continue to flow until scale $h_2$ (possibly $h_2=-\infty$).

The multi-scale procedure is well defined, and the effective potentials are, step by step, given by {\it convergent} expansions, provided: (i) $U$ is small enough, (ii) $\xi_{\o,h}$ remain small for all scales, and (iii) $Z_{\r,\o,h}, v_{\o,h}$ remain close to their initial (bare) values, for all scales. 
Note that, in order for condition (ii) to be valid, we need to properly 
fix the initial condition on the chemical potential, as discussed in the following. In addition, note that, once that the flows of $Z_{\r,\o,h}$ and $v_{\o,h}$ are controlled, then the marginal contributions to the effective 
source term $\LL B^{(h)}(\psi,0,A)$ are automatically under control, thanks to \eqref{eq:lbh} and following lines.

The key fact, which allows us to control the flow of the effective couplings, is that, since the electron-electron interaction is {\it irrelevant}, with scaling dimension $D=-1$ (cf. with \eqref{eq:scdim}),
then the scaling dimensions of all diagrams with at least one interaction vertex can be effectively improved by one, see Ref.\cite{GM}. In particular, 
$|\b^\xi_{\o,h}|\le c_\e|U|2^{(1-\e) h}$, for any $\e>0$ and a suitable constant 
$c_\e>0$, and similarly for the beta functions of $Z_{\r,\o,h}$ and $v_{\o,h}$. [The reason why we lose, in general, an $\e$ in the decay exponent as $h\to-\infty$, is that we need to use a 
little bit of decay $2^{\e h}$ in order to sum over all diagrams and scales, see Ref.\cite{GM} for details.]

In order to guarantee that the flow of the chemical potential remains bounded, we fix the initial data (via a fixed point theorem, such as the contraction mapping theorem) so that 
$\lim_{h\to-\infty}\xi_{\o_2,h}=0$, in the limit as $h_2\to-\infty$. Thanks to the dimensional gain of $2^{(1-\e)h}$, due to the irrelevance of the interaction, we actually find that $\xi_{\o_2,h}$ tends to zero, 
as $h\to-\infty$, exponentially fast: $|\xi_{\o_1,h}|\le$(const.)$|U|2^{(1-\e)h}$. Once we imposed that $\xi_{\o_2,h}$ remains bounded for all scales $h\le 0$, we can a posteriori check that $\xi_{\o_1,h}$ is 
also bounded for all scales $h_1\le h\le 0$: 
in fact, the beta function $\b^\xi_{\o_1,h}$, for $h\ge h_1$, can be rewritten as $\b^\xi_{\o_2,h}+[\b^\xi_{\o_1,h}-\b^\xi_{\o_2,h}]$, where the difference in square brackets can be straightforwardly shown to be proportional 
to $m_{\o_1}$ [if all the masses $m_{\o,h}$ were zero, then the model would be symmetric under the exchange of $\o$ in $-\o$, as in Ref.\cite{GM}, see also Section \ref{sec.symm} above; therefore, the difference $\b^\xi_{\o_1,h}-\b^\xi_{\o_2,h}$ between the 
contributions with different valley indices must be proportional to a mass term $|m_{\o,h}|$, which is smaller than (const.)$|m_{\o_1}|$]. Therefore, the flow of $\xi_{\o_1,h}$, for $h\ge h_1$, remains close to 
the one of $\xi_{\o_2,h}$ (which is uniformly bounded for all scales), up to terms that are proportional to $m_{\o_1}$ and, therefore, are bounded by (const.)$|U||m_{\o_1}|2^{-h}2^{(1-\e) h}$ (here $2^{-h}$ is the 
dimensional amplification factor arising from the scaling dimension $D=+1$ of the chemical potential terms, while $2^{(1-\e)h}$ is the dimensional gain coming from the irrelevance of the interaction).
Recalling that $2^{h_1}\simeq |m_{\o_1}|$, we find that $|\xi_{\o_1,h}|\le$(const.)$|U|2^{(1-\e)h}$, for all 
scales $h\ge h_1$.

Finally, once the chemical potential is fixed so that $|\xi_{\o,h}|\le$(const.)$|U|2^{(1-\e)h}$, we immediately 
infer that the beta functions of $Z_{\r,\o,h}$ and $v_{\o,h}$ are bounded by (const.)$|U|2^{(1-\e)h}$, as well: therefore, their flows converge exponentially fast, and the 
dressed values of $Z_{\r,\o,h}$ and $v_{\o,h}$ are analytic functions of $U$, analytically close to their bare values.

\subsection{Lowest order computations}

The discussion in the previous section guarantees that, once the chemical potential is properly fixed, then the flows of the chemical potential, wave function renormalizations, and Fermi velocity converge 
exponentially fast. The values of the chemical potential, as well as of the dressed wave functional renormalizations, dressed Fermi velocity, and dressed critical lines are expressed in terms of 
convergent expansions (they are {\it analytic} functions of $U$), which are dominated by the first non trivial order in perturbation theory, provided $U$ is not too large (note that the condition of convergence of the renormalized expansion is {\it uniform} in the gap, and is valid, in particular, {\it on the critical line}).
The explicit lowest order contributions to the chemical potential $\xi$, to the renormalized Fermi velocity $v_R\equiv v_{\o_2,-\infty}$ and the wave function renormalizations $Z_{\r,R} \equiv Z_{\r,\o_2,-\infty}$
on the renormalized critical line $h_2=-\infty$ are the following: 
\begin{enumerate}
\item Chemical potential:
\be
\xi=-\frac{U^2}{2}\sum_{\r=1}^2\int \frac{d\kk d\qq}{(2\p|\BBB|)^2}\hat g_{\r\r}(\kk+\pp_F^{\o_2})\hat g_{\r\r}(\qq)\hat g_{\r\r}(\kk+\qq);\nn\ee
\item Fermi velocity: 
\be
v_R=\frac32 t_1-iU^2\int \frac{d\kk d\qq}{(2\p|\BBB|)^2}\partial_{k_1}\hat g_{12}(\kk+\pp_F^{\o_2})\hat g_{12}(\qq)\hat g_{21}(\kk+\qq);\label{eq:vR}\ee

\item Wave function renormalizations:
\be
Z_{\r,R}=1+iU^2\int \frac{d\kk d\qq}{(2\p|\BBB|)^2}\partial_{k_0}\hat g_{\r\r}(\kk+\pp_F^{\o_2})\hat g_{\r\r}(\qq)\hat g_{\r\r}(\kk+\qq).\label{eq:ZrR}\ee
\end{enumerate}

Moreover, the equation for the critical line $h_2=-\infty$ reads: 
\be m_{\o_2}=\frac{U}2\int\frac{d\vec k}{|\BBB|}\frac{m(\vec k)}{\sqrt{m^2(\vec k)+t_1^2|\O(\vec k)|^2}}\;,\nn\ee
where $m_\o$, $m(\vec k)$ and $\O(\vec k)$ where defined after \eqref{hbloch}. This is a fixed point equation for $m_{\o_2}$, whose solution leads to the plot in Fig.\ref{figaltezza}.

Note that, as discussed in Sect.\ref{sec.symm}, there is no symmetry reason why $Z_{1,R}$ should be equal to $Z_{2,R}$. Actually, an explicit computation shows that $Z_{1,R}-Z_{2,R}$ is different from zero
along the critical line, unless we are at one of the highly symmetric points $\phi=0$ or $\phi=\p/2$, see Fig.\ref{fig1-2}, where we plot the value of $Z_{1,R}-Z_{2,R}$ on the critical line at second order in $U$, for two different values of $U$. 

\section{Quantization of the conductivity}\label{sec4}

In this section we compute the jump discontinuity of the Hall conductivity across the critical line, as well as the value of the longitudinal conductivity 
on the same line, and prove a universality result for both of them, i.e., we prove that their values are quantized and exactly independent of the 
interaction strength $U$. Note that this fact is highly non-trivial, due to the unusual renormalization of the Fermi velocity and of the wave function renormalizations,
which depends explicitly on the spinor index and break the asymptotic relativistic invariance of the propagator: the cancellations behind universality 
need to take lattice (and, therefore, RG-irrelevant) effects into account, and do not follow from asymptotic relativistic computations. 

We stress that our result is exact at all orders of the (convergent, renormalized) expansion for the conductivity. One key ingredient used in the proof is 
the lattice Ward Identity \eqref{1.14}, which is rigorously valid (without any sub-leading correction), thanks to the exact lattice symmetries and the fact that the correlations 
appearing at both sides can be computed in terms of convergent expansions, following from the multi-scale construction described above. 

\subsection{Quantization of the Hall conductivity across the critical line}

Here we compute the universal jump discontinuity of the Hall conductivity across the renormalized critical line. For the moment, we assume not to be at the graphene points $W\,,\phi=0$ and $W=0$, $\phi = \pi$; we shall discuss later the (straightforward) adaptation to these special cases. Therefore, the goal is to compute:
\be
\D = \lim_{m_R\to 0^{+}} \s_{12} - \lim_{m_R\to 0^{-}}\s_{12}\;,\nn
\ee
where $m_R\equiv m_{\o_2,h_2}$ is the mass gap of the dressed propagator. The condition that we are not at a graphene point means that 
$m_{\o_1,h_1}$ should be kept finite as $m_R\to 0$.
Using the definition \eqref{eq:sijdef}, as well as the fact that $\hat K_{ij}(\pp)$ is differentiable in $\pp$ outside the critical line, we can rewrite  
$$\D = -\frac1{A}\Big[\lim_{m_{R} \to 0^{+}} \partial_{p_0}\hat K_{12}({\bf 0}) - \lim_{m_{R}\to 0^{-}}\partial_{p_0}\hat K_{12}({\bf 0})\Big].$$
The interacting current-current correlation can be computed via the multiscale renormalized expansion discussed in Sect. \ref{sec.RG}: in particular, proceeding as in Ref.\cite{GMP},
among the contributions to $\hat K_{ij}$ we can distinguish the dominant contribution, coming from the `dressed bubble', from the sub-dominant one, which is the sum over all the
renormalized diagrams with at least one interaction term. Thanks to the irrelevance of the interaction, these sub-dominant diagrams have a dimensional gain (of order $2^h$ on scale $h$),
which makes the corresponding contribution to $\hat K_{ij}(\pp)$ {\it differentiable} at $\pp={\bf 0}$, in the limit $m_R\to 0$. In particular, they give zero contribution to $\D$.

The dominant contribution to $\hat K_{ij}(\pp)$ (i.e., the `dressed bubble') is
\bea \hat K_{ij}^{\text{dom}}(\pp)&=&-2\int \frac{d\kk}{2\p|\BBB|}{\rm Tr}\big\{\hat S_2(\kk)\hat \G_{i}(\kk,\pp)\times\nn\\
&& \times\, \hat S_2(\kk+\pp)\hat \G_{j}(\kk+\pp,-\pp)\big\},\nn\eea
where $\hat \G_{j}$ is the vertex function defined in \eqref{eq:vf}, and the factor 2 in front of the integral takes into account the spin degrees of freedom. 
Both $\hat S_2(\kk)$ and $\hat \G_i(\kk,\pp)$ are given by convergent renormalized series, which depend on the details of the microscopic model.

The finite contribution to the jump-discontinuity of $\partial_{p_0}\hat K_{12}({\bf 0})$ across $m_R=0$ comes from the integration over $\kk$ in the vicinity of $\pp_F^{\o_2}$, since the rest is continuous 
as $m_R\to 0$. For the same reason, for the purpose of computing $\D$, we can replace $\hat\G_{i}(\kk,\pp)$ by $\hat\G_{i}(\pp_F^{\o_2})=\hat\G_{i}(\pp_F^{\o_2},{\bf 0})$, and 
$\hat S_2(\kk)$ by its linearization $\bar S(\kk')$ at $\pp_F^{\o_2}$,  
\be
\bar S(\kk') = \begin{pmatrix} -ik_0 Z_{1,R}+m_R & -v_R(-ik_1'+\o_2 k_2')\\
-v_R(ik_1'+\o_2 k_2') & -ik_0 Z_{2,R}-m_R\end{pmatrix}^{-1}\;,\label{eq:linprop}
\ee
where $Z_{\r,R}$ and $v_R$ are {\it analytic} functions of $U$, for $U$ small, whose expansions at second order in $U$ are given explicitly by \eqref{eq:vR}-\eqref{eq:ZrR}.
Recall that, a priori, $\hat\G_{i}(\pp_F^{\o_2})$ are complicated infinite series in $U$. Thus, a direct computation of the jump-discontinuity, starting from the expression of the dressed bubble and 
from the Feynman rules for the generic term in the renormalized expansions for $Z_{\r,R}$, $v_R$ and $\hat\G(\pp_F^\o)$, would be hopeless. 

The key fact is that, thanks to the Ward 
Identity \eqref{eq:wi-2}, 
\be\label{eq:WIvertex}
\hat\G_{i}(\pp_F^{\o_2})=\partial_{k_i'}\bar S^{-1}({\bf 0})
\ee
that is, 
\be\label{eq:WIvertex2}
\hat\G_{1}(\pp_F^{\o_2})=-v_R\s_2\;,\quad \hat\G_{2}(\pp_F^{\o_2})=-\o_2v_R\s_1\;.
\ee
Therefore,
\bea &&\D=\Big(\lim_{m_R\to 0^+}-\lim_{m_R\to 0^-}\Big)\int_{|\vec k'|\le \e} \frac{d\vec k'}{2\p^2}\int_{\mathbb R}\frac{dk_0}{2\p}\times\nn\\
&&\times{\rm Tr}\big\{ \bar S(\kk') \partial_1\bar S^{-1}({\bf 0}) \partial_0\bar S(\kk')\partial_2\bar S^{-1}({\bf 0})\big\},\label{eq:Delta}\eea
where we used that $A|\BBB|=4\p^2$, and we denoted by $\e$ a small, arbitrary, positive constant. 
Using the identity 
\be
\partial_0 \bar S(\kk') \bar S^{-1}(\kk')=-\bar S(\kk') \partial_0\bar S^{-1}(\kk'),
\ee
and replacing $\bar S(\kk') \partial_0\bar S^{-1}(\kk')$ by 
$\bar S(\kk') \partial_0\bar S^{-1}({\bf 0})$ (which is allowed, for the purpose  of computing $\D$, simply because the difference is continuous at $m_R=0$),
we can further rewrite $\Delta$ as 
\bea &&\D=-\Big(\lim_{m_R\to 0^+}-\lim_{m_R\to 0^-}\Big)\int_{|\vec k'|\le \e} \frac{d\vec k'}{2\p^2}\int_{\mathbb R}\frac{dk_0}{2\p}\times\nn\\
&&\times{\rm Tr}\big\{ \bar S(\kk') \partial_1\bar S^{-1}({\bf 0}) \bar S(\kk')\partial_0\bar S^{-1}({\bf 0})\bar S(\kk')\partial_2\bar S^{-1}({\bf 0})\big\}.\nn\eea
The integral over $k_0$ can be evaluated explicitly and, after a straightforward computation, we get 
\bea &&\D=\frac{\o_2v_R^2}{4\p^2}\frac{Z_{1,R}+Z_{2,R}}{(Z_{1,R}Z_{2,R})^2}\lim_{m_R\to 0^+}m_R\times\nn\\
&&\times \int_{|\vec k'|\le \e} d\vec k'\Big[\frac{m_R^2}{4}\big(\frac1{Z_{1,R}}+\frac1{Z_{2,R}}\big)^2+\frac{v_R^2|\vec k'|^2}{Z_{1,R}Z_{2,R}}\Big]^{-3/2}.\nn\eea
Thus, introducing
\be\label{eq:vmtilde}
\widetilde v_{R} = \frac{v_{R}}{\sqrt{Z_{1,R}Z_{2,R}}}\;,\quad \widetilde m_{R} = m_{R} \frac{Z_{1,R} + Z_{2,R}}{Z_{1,R}Z_{2,R}}\;,
\ee
we see that $\D$ can be rewritten as, performing the change of variables $\widetilde v_{R}\vec k'\to \vec k'$:
\bea
\D &=& \frac{\o_{2}}{4\pi^2} \lim_{\widetilde m_{R}\to 0^{+}} \widetilde m_{R} \int_{|\vec k'|\le \e \widetilde v_{R}} d\vec k'\Big[\frac{\widetilde m_R^2}{4} + |\vec k'|^2\Big]^{-3/2}\;\nn\\
&=& \frac{\o_{2}}{4\pi^2} \lim_{\widetilde m_{R}\to 0^{+}} \int_{|\vec k'|\le \e \widetilde v_{R} / \widetilde m_{R}} d\vec k'\Big[\frac{1}{4} + |\vec k'|^2\Big]^{-3/2}\nn\\
&=& \frac{\o_{2}}{\pi}\;,
\eea
where we recall that the result is expressed in units such that $e^{2} = \hbar = 1$. 
Therefore, the cancellation between the parameters $v_R$, $Z_{1,R}$, $Z_{2,R}$ gives a universal result. Finally, at the graphene points, the analogous computation gives twice the same value, 
because of an extra factor 2 coming from the valley degeneracy.

\subsection{Quantization of the longitudinal conductivity on the critical line}

A similar discussion as the one in the previous subsection can be repeated for the {\it longitudinal} conductivity on the renormalized critical line. The point here, as compared to the computation of $\D$ in the previous subsection, is to take first the limit $m_R\to 0$, and then $p_0\to 0^+$ (recall the definition of conductivity, Eq. \eqref{eq:sijdef}). Once again, we assume for definiteness not to be exactly at the graphene point (a similar discussion applies there, too). 

Note that, by the very definition of current-current correlations, $\hat K_{ii}(p_0,\vec 0)$ is {\it even} in $p_0$. Therefore, all the contributions to $\hat K_{ii}(p_0,\vec 0)$ that are differentiable in $p_0$ give zero 
contribution to the longitudinal conductivity on the critical line. By repeating a strategy analogous to the one that led us to \eqref{eq:Delta}, for the purpose of computing the longitudinal conductivity on the 
critical line, we can: (i) replace the full current-current correlation by its dominant contribution (from the `dressed bubble'); (ii) restrict the integration over the loop  momenta in the vicinity of $\pp_F^{\o_2}$; (iii)
linearize the propagators and vertex functions around $\pp_F^{\o_2}$; (iv) use the Ward identity Eq. (\ref{eq:WIvertex}) to replace the vertex functions by the derivatives of the inverse two-point function. 

After these replacements, we get (denoting the value of the longitudinal conductivity on the critical line by $\s_{ii}^{\text{cr}}$):
$$\s_{ii}^{\text{cr}} = \frac2{A}\lim_{p_0\to 0^+}\frac1{p_0}\int_{|\vec k'|\le \e} \frac{d\vec k'}{|\BBB|}\int_{\mathbb R}\frac{dk_0}{2\p} \big[F(\kk',p_0)-F(\kk',0)\big],$$
with
\bea 
F(\kk',p_0)&=&\Tr\big\{ \bar S_0(\kk') \partial_i \bar S_0^{-1}({\bf 0}) \bar S_0(k_0+p_0,\vec k')\partial_i\bar S_0^{-1}({\bf 0})\big\}\nn\\
&=&v_{R}^{2}\Tr\big\{ \bar S_0(\kk') \s_{i} \bar S_0(k_0+p_0,\vec k') \s_{i} \big\}
\eea
where $\bar S_0(\kk')$ is the linearized propagator \eqref{eq:linprop}, computed at $m_R=0$, and the last step follows from (\ref{eq:WIvertex}), (\ref{eq:WIvertex2}). By evaluating the integral over $k_0$ 
explicitly, and setting $\widetilde v_{R}=v_R/\sqrt{Z_{1,R}Z_{2,R}}$ as in Eq. (\ref{eq:vmtilde}), the computation of $\s_{ii}^{\text{cr}}$ reduces to the contribution of just one Dirac cone to the longitudinal 
conductivity of noninteracting graphene \cite{SPG, GMP}, with Fermi velocity $\widetilde v_{R}$. Thus, proceeding as in Ref.\cite{GMP}, we get, in units such that $e^{2} = \hbar = 1$:
\be
\s_{ii}^{\text{cr}} = \frac1{2\p}\lim_{p_0\to 0^+}\int_0^{\widetilde v_{R}\e}  \frac{p_0}{p_0^2+4x^2}\, dx=\frac18.
\ee
Notice that, as for graphene, the Fermi velocity (in general a nontrivial function of the Hubbard interaction strength $U$) disappears, thus yielding a universal result. The analogous computation performed at the 
graphene points gives twice the same value, in agreement with the result of Ref.\cite{GMP}.

\section{Conclusions}

We studied the Haldane-Hubbard model by rigorous Renormalization Group techniques.
Our analysis predicts that the critical lines separating the distinct topological phases are modified non-trivially by the Hubbard interaction, in particular that the non-trivial topological phase, characterized by the topological quantum number $\n=\pm2$, is enlarged by weak repulsive interactions. Moreover, our results rule out the presence of new interaction-induced topological phases in the vicinity of the phase boundaries. Such 
predictions may be verified experimentally in optical lattice realizations of the system\cite{E}, where the on-site interaction can be produced and tuned by means of Feshbach resonances. Concerning numerical 
simulations, our results agree with those of Ref.\cite{Van, Hub}. 

The interaction affects the relativistic structure of the two-point function by non-universal renormalizatized coefficients, which differ from those obtained by 
approximate treatments of the system based on the effective Dirac theory. In particular, we find that there are two different wave function renormalizations, one for each pseudo-spin index. 
Despite the non-universal renormalization of the two-point function and of the vertex functions, lattice Ward identities guarantee the quantization and the universality of the conductivity matrix at the critical line. 
Concerning the transverse conductivity $\s_{12}$, its quantization follows from topological arguments; however, these arguments do not provide any information regarding which values $\s_{12}$ might take. For 
instance, numerical and mean-field analyses predict that, at intermediate coupling strengths, new topological phases might appear, corresponding to the values $\s_{12} = \pm e^{2}/h$, which are not present in the 
noninteracting theory. Our exact analysis rules out such new phases at small coupling.

The second part of our result focuses on the critical longitudinal conductivity $\s^{\text{cr}}_{11}$ (away from criticality $\s_{11}$ is trivially zero). In constrast to $\s_{12}$, this quantity is not protected by any 
topological argument. Nevertheless, we show that it is universal: all interaction and lattice corrections disappear. Each Dirac cone contributes with a universal quantum of conductivity $(e^{2}/h)(\pi/4)$; in 
particular, at the doubly critical points where the two critical curves cross (see Fig. \ref{figaltezza}), the critical longitudinal conductivity is $(e^{2}/h)(\pi/2)$, which is the same value measured in graphene \cite{Nair}.

Our results require the interaction to be weak and short-range; instead, different features are expected in the presence of long-range interactions.
For instance, it is known that, at the graphene point, long-range interactions have dramatic effects on several physical properties \cite{GGV, GMPgauge}, and their role on the 
renormalization of the optical conductivity is still actively debated
\cite{L1,L11,L12,L13,L14,L15,L16,L17,L18,L19}. We expect such effects to have profound implications for the Haldane-Hubbard model, especially in the proximity of the critical lines separating the different 
topological phases. We plan to investigate this issue in future work.

\bigskip

The work of A.G. has been carried out thanks to the support of the
A*MIDEX project Hypathie (no ANR-11-IDEX-0001-02) funded by the ``Investissements d'Avenir''
25 French Government program, managed by the French National Research Agency (ANR), and by a C.N.R.S. visiting professorship
spent at the University of Lyon-1.
The work of M.P. has been carried out thanks to the support of the NCCR SwissMap.

\appendix

\section{Peierls' substitution and the bare vertex functions}\label{app.ver}

In order to define the current, we couple the electron gas to an external vector potential $\vec A$,
by multiplying the hopping strength from $\vec y$ to $\vec x$ by an extra phase factor $e^{i(\vec y-\vec x)\int_0^1 \vec A((1-s)\vec x+s\vec y)ds}$
(Peierls' substitution). We denote by $H(\vec A)$ the modified Hamiltonian, and let the (paramagnetic) current be 
$J_{\vec p,i}=\d H(\vec A)/\d \hat A_{\vec p,i}\big|_{\vec A=\vec 0}$, where $i=1,2$ label the two (orthogonal) coordinate directions $\hat e_1=(1,0)$ and $\hat e_2=(0,1)$.
An explicit computation leads to \eqref{eq:curr}, with 
$\vec M(\vec k,\vec p)=\begin{pmatrix}\vec M_{11}(\vec k,\vec p)&\vec M_{12}(\vec k,\vec p)\\
\vec M_{21}(\vec k,\vec p)&\vec M_{22}(\vec k,\vec p)\end{pmatrix}$ and, defining $\h_{x}=(e^{-ix}-1)/(-ix)$, 
\bea &&\vec M_{11}(\vec k,\vec p)
=-it_2\sum_{j=1}^3\sum_{\a=\pm}
\a\vec\g_j\h_{\a\vec p\cdot\vec\g_j}e^{i\a(\phi-\vec k\cdot\vec \g_j)},\nn\\
&&\vec M_{12}(\vec k,\vec p)
=-it_1\sum_{j=1}^3\vec\d_j\h_{\vec p\cdot \vec\d_j}e^{-i\vec k(\vec \d_j-\vec \d_1)}\;,\nn\\
\nn\eea
$\vec M_{21}(\vec k,\vec p)=-\vec M_{12}(-\vec k-\vec p,\vec p)$ and 
$\vec M_{22}(\vec k,\vec p)=-e^{-i\vec p\cdot\vec \d_1}\vec M_{11}(-\vec k,-\vec p)$. 

\section{Details of the numerical computations}\label{app.numerics}
In this appendix, we discuss some of the details of the numerical computations from which Figs.\ref{figaltezza}-\ref{fig1-2} were produced. The program used to carry them out is available online \cite{hhtop}, has been named {\tt hhtop}, and is released under an Apache license. The source code includes a documentation file, in which the computations are described in greater detail.

\subsection{Integration scheme}\label{app.gauss_legendre}
The numerical computations carried out in this work involve numerical evaluations of integrals. The algorithm that was used to carry these out is based on {\it Gauss-Legendre} quadratures, by which, given an integer $N>1$, an integral is approximated by a discrete sum with $N$ terms:
\begin{equation}
  \int_{-1}^1dx\ f(x)=\sum_{i=1}^Nw_if(x_i)+\mathfrak R_N
\end{equation}
where $x_1<\cdots<x_N$ are the roots of the $N$-th Legendre polynomial $P_N$, and
\begin{equation}
  w_i:=\frac 2{(1-x_i^2)P_N'(x_i)}.
\end{equation}

If $f$ is an analytic function, then one can show that the remainder $\mathfrak R_N$ decays exponentially in $N$. However, in order to compute the difference of the wave-function renormalizations, we need to compute the integral of an integrand that, instead of being analytic, is a class-2 Gevrey function (a class-$s$ Gevrey function is a $\mathcal C^\infty$ function whose $n$-th derivative is bounded by $(\mathrm{const}.)^n(n!)^s$, so that analytic functions are class-1 Gevrey functions). The remainder $\mathfrak R_N$ can be shown to be bounded, if $f$ is a class-$s$ Gevrey function with $s\ge1$ and $N$ is large enough (independently of $f$ and $s$), by
\begin{equation}
  |\mathfrak R_N|\leqslant c_0c_1^{s-1}(2N)^{1-\frac1s}e^{-b(2N)^{\frac1s}}s!
\end{equation}
for some $c_0,c_1,b>0$, that only depend on $f$. For a proof of this statement, see lemma~\-A3.1 in the documentation of {\tt hhtop}\cite{hhtop}. In short, this estimate is obtained by expanding $f$ in Chebyshev polynomials, and using a theorem of A.C. Curtis and P. Rabinowitz\cite{CR72} that shows that, if $f$ is the $j$-th Chebyshev polynomial, then $\mathfrak R_N$ is bounded uniformly in $j$. The decay of the coefficients of the Chebyshev expansion of class-$s$ Gevrey polynomials allows us to conclude.

\subsection{First-order renormalization of the critical line}\label{app.critical_line}
At first order in $U$, the correction $F_{\pm,R}(U,W,\phi)$ appearing in (\ref{fpm}) is
\begin{equation}
  F_{\pm}=\frac U2\int_{\mathcal B}\frac{d\vec k}{|\mathcal B|}\ \frac{m(\vec k)}{\sqrt{m^2(\vec k)+t_1^2|\Omega(\vec k)|^2}}.
  \label{eq.Fnumeric}
\end{equation}

There is a single, minor, pitfall in the numerical evaluation of $F_{\pm}$: we wish to use Gauss-Legendre quadratures (see App.\ref{app.gauss_legendre}) to carry out the computation, but the integrand in (\ref{eq.Fnumeric}) is not smooth: indeed, if $W=\pm3\sqrt3t_2\sin\phi$, then its second derivative diverges at $\vec p_F^\pm$ due to the divergence of the derivative of $\sqrt{\cdot}$. However, by switching to polar coordinates $\vec k=p_F^\pm+\rho(\cos\theta,\sin\theta)$, this singularities is regularized, that is, the integrand becomes a smooth function of $\rho$ and $\theta$. At this point, there is yet another danger to avoid: while the integrand is smooth, the upper bound of the integral over $\rho$ is a function of $\theta$, which is, due to the rhombic shape of $\mathcal B$, only smooth by parts. The integral over $\theta$ must, therefore, be split into parts in which the bounds of the integral over $\rho$ are smooth. This can be done very easily using the $\frac{2\pi}3$ rotation symmetry. Once both of these traps have been thwarted, Gauss-Legendre quadratures yield very accurate results.

In order to compute the correction to the critical line, we solve
\begin{equation}
  W\pm3\sqrt3t_2\sin\phi-F_{\pm}(\phi,W)=0
  \label{eq.mass1}
\end{equation}
for $W$ and $\phi$. For the sake of clarity, we have made the $(\phi,W)$ dependence of $F_{\pm}$ explicit. To solve (\ref{eq.mass1}), we fix $\phi$, and use a Newton algorithm to compute the critical value of $W$: we set $W_0=\mp3\sqrt3t_2\sin\phi$, and compute
\begin{equation}
  W_{n+1}=W_n-\frac{W_n\pm3\sqrt3t_2\sin\phi-F_{\pm}(\phi,W_n)}{1-\partial_WF_{\pm}(\phi,W_n)}.
\end{equation}
Provided $W_0$ is not too far from the solution of (\ref{eq.mass1}), $W_n$ converges {\it quadratically} (i.e. $|W_{n+1}-W_n|\le (\mathrm{const.})|W_n-W_{n-1}|^2$, in which the constant depends on the supremum of $\partial^2_WF_{\pm}$, which is bounded) to the solution of (\ref{eq.mass1}).

\subsection{Second-order wave function renormalization}
At second order in $U$, $Z_{1,R}-Z_{2,R}$ is
\begin{equation}
  U^2(z_1-z_2)=U^2 i\left(\partial_{k_0}s_1|_{k_0=0}-\partial_{k_0}s_2|_{k_0=0}\right)
\end{equation}
where
\begin{equation}
  s_i:=\int_{\mathcal B}\frac{d\vec pd\vec q}{|\mathcal B|^2}\int_{-\infty}^\infty\frac{dp_0dq_0}{(2\pi)^2}
  \hat g_{i,i}(\mathbf p)\hat g_{i,i}(\mathbf q)\hat g_{i,i}(\mathbf p+\mathbf q-\mathbf k_F^\omega).
\end{equation}
The computation is carried out on the critical line, that is, when $W=-\omega3\sqrt3t_2\sin\phi$. The integrals over $p_0$ and $q_0$ can be carried out explicitly:
\begin{equation}
  \begin{array}{>\displaystyle l}
    z_1-z_2
    =\int_{\mathcal B}\frac{d\vec pd\vec q}{|\mathcal B|^2}
    \cdot\\[0.3cm]\hskip15pt\cdot
    \left(\frac{(\xi_p+\xi_q+\xi_F)(\frac{m_p}{\xi_p}+\frac{m_q}{\xi_q}-\frac{m_F}{\xi_F}-\frac{m_pm_qm_F}{\xi_p\xi_q\xi_F})Z}{(Z^2-(\xi_p+\xi_q+\xi_F)^2)^2}\right)
  \end{array}
  \label{eq.z1-z2}
\end{equation}
where, using the definitions of $m(\vec k)$, $R(\vec k)$ and $\Omega(\vec k)$ after \eqref{hbloch} and after \eqref{prop}, $m_p\equiv m(\vec p)$, $m_q\equiv m(\vec q)$, $m_F\equiv m(\vec p+\vec q-\vec p_F^\omega)$, $\xi(\vec k):=\sqrt{m(\vec k)+t_1^2|\Omega(\vec k)|^2}$, $Z:=R(\vec p)+R(\vec q)-R(\vec p+\vec q-\vec p_F^\omega)$ and $\xi_p\equiv \xi(\vec p)$, $\xi_q\equiv \xi(\vec q)$, $\xi_F\equiv \xi(\vec p+\vec q-\vec p_F^\omega)$.

The numerical evaluation of the integral in (\ref{eq.z1-z2}) involves a similar difficulty to that in (\ref{eq.Fnumeric}): the integrand has divergent derivatives if any of the following conditions hold: $\vec p=\vec p_F^\omega$, $\vec q=\vec p_F^\omega$ or $\vec p+\vec q=2\vec p_F^\omega$. These singularities cannot be regularized by changing $\vec p$ and $\vec q$ to polar coordinates, since $\xi_F$ is a singular function of the polar coordinates of $\vec p$ and $\vec q$ (due to the fact that it behaves, asymptotically, as $\vec p-\vec q$ approaches $2\vec p_F^\omega$, as $|\vec p+\vec q-2\vec p_F^\omega|$, which has divergent second derivatives). However, there are coordinates, which we call {\it sunrise coordinates} (since $s_i$ is the value of the so-called {\it sunrise} Feynman diagram), which regularize these singularities. Their expression is rather long, and will not be expounded here; the interested reader is invited to consult the documentation file bundled with the source code of {\tt hhtop}\cite{hhtop}. 
Once written in terms of the sunrise coordinates, the integral in (\ref{eq.z1-z2}) can be computed using Gauss-Legendre quadratures very accurately.


\begin{thebibliography}{9}
\bibitem{brillouin} In formul\ae, $\BBB=\{\vec k: \vec k=a_1\vec\ell^*_1+a_2\vec\ell^*_2,$ $a_i\in[0,1)\}$,
where $\vec\ell^*_{1,2}=\frac{2\pi}{3}(1,\mp\sqrt{3})$ are the basis vectors of the reciprocal lattice $\L_A^*$.
\end{thebibliography}
\end{document}